\documentclass[fleqn,usenatbib]{mnras}

\usepackage[T1]{fontenc}
\usepackage{ae,aecompl}

\usepackage{graphicx}	
\usepackage{amsmath}	
\usepackage{amssymb}	

\usepackage{csquotes}

\usepackage{etoolbox}
\makeatletter
\patchcmd\@combinedblfloats{\box\@outputbox}{\unvbox\@outputbox}{}{\errmessage{\noexpand patch failed}}
\makeatother

\newcommand{\mbh}{M_{\rm BH}}
\newcommand{\msun}{{\rm M}_{\sun}}
\newcommand{\ledd}{L_{{\rm Edd}}}

\newcommand{\ergs}{{\rm \,erg\,s^{-1}}}

\newcommand{\lx}{L_{\rm X}}
\newcommand{\lr}{L_{\rm R}}

\title[FRED lightcurve in NGC 7213]{A decades-long fast-rise-exponential-decay flare in low-luminosity AGN NGC 7213}

\author[Z. Yan \& F. G. Xie]{Zhen Yan$^1$\thanks{E-mail: zyan@shao.ac.cn (ZY); fgxie@shao.ac.cn (FGX)}, Fu-Guo Xie$^1$$^\star$\\
	$^1$Key Laboratory for Research in Galaxies and Cosmology, Shanghai Astronomical Observatory, \\
	Chinese Academy of Sciences, 80 Nandan Road, Shanghai 200030, China.}

\date{Accepted .... Received ...; in original form ...}
\pubyear{2017}

\begin{document}
	
\pagerange{\pageref{firstpage}--\pageref{lastpage}}
\maketitle

\label{firstpage}

\begin{abstract}
We analysed the four-decades-long X-ray light curve of the low-luminosity active galactic nucleus (LLAGN) NGC 7213 and discovered a fast-rise-exponential-decay (FRED) pattern, i.e. the X-ray luminosity increased by a factor of $\approx 4$ within 200d, and then decreased exponentially with an $e$-folding time $\approx 8116$d ($\approx 22.2$ yr). For the theoretical understanding of the observations, we examined three variability models proposed in the literature: the thermal-viscous disc instability model, the radiation pressure instability model, and the tidal disruption event (TDE) model. We find that a delayed tidal disruption of a main-sequence star is most favourable; either the thermal-viscous disk instability model or radiation pressure instability model fails to explain some key properties observed, thus we argue them unlikely.
\end{abstract}
	
\begin{keywords}
{accretion, accretion disks --- black hole physics  --- galaxies: individual: NGC 7213--- galaxies: nuclei}
\end{keywords}

\section{Introduction}
\label{intro}

Active galactic nuclei (AGN) have been found to be variable (in flux and/or spectrum) at all wavebands, from radio to X-rays, and even $\gamma$-rays, with variability amplitude being as large as a factor of $\sim 100$ \citep{Ulrich1997, Peterson2001, Netzer2008}. Moreover, the AGN are also known to be variable on different time-scales, i.e. from hours to years or even decades/centuries. Considering the lack of sufficient spatial resolution in most existing observations, variability studies have been essential to diagnose the physics, structure, and kinematics of the central regions of AGN. The time-scales, the spectral changes, and the correlations and delays between variations in the different continuum or line components will reveal the nature and location of different physical components and on their interdependencies. For example, in the reverberation mapping method, the time lag between optical/ultraviolet continuum and H$\beta$ (Mg{\sc ii}, C{\sc iv} also) emission lines is crucial to constrain the size of broad line regions in AGN (for a recent review, see \citealt{Bentz2016}).
	
Theoretically speaking, the variability of AGN on different time-scales is believed to be triggered by different mechanisms. Over short time-scale of hours to months, the leading variability mechanism is the fluctuation within the accretion disc (e.g., \citealt{McHardy2004, Papadakis2004, Uttley2005, Done2005, Arevalo2006, McHardy2006, Netzer2008, Kelly2009, Kelly2011, McHardy2010}). Additionally, the variability in infrared, optical and soft X-rays on this time-scale also possibly relates to either the change in illuminating X-ray emission \citep[e.g., ][]{Shappee2014, Denney2014}, or in the dynamics of absorbing gas \citep{Grupe2013,LaMassa2015,Runnoe2016}. Besides, in Blazars, their variability may be caused by the shock and/or turbulence within the jet. Dramatic (large amplitude) flux variation on time-scales over years has been clearly observed in numerous AGN (e.g. \citealt{Strotjohann2016}), and several mechanisms are proposed (e.g., \citealt{Valtonen2008, Elitzur2014, Merloni2015}). Among them, one interesting scenario is the tidal disruption event (TDE) model, i.e. considering a star to be captured and tidally disrupted by the super-massive black hole (SMBH) of a galaxy (\citealt{Rees1988}, see \autoref{sec:tde}). The TDE usually produces an outburst (e.g., \citealt{Drake2011, Blanchard2017a}). On longer time-scales of $\sim 10^{3-6}$ yr, the variability can be triggered by the instabilities in the accretion discs (e.g., the radiation-pressure instability, see \citealt{Janiuk2004, Janiuk2011, Czerny2009}). Over cosmic time, consensus has been reached that during the non-AGN phase gases will accumulate in a quiescent disc around the SMBH located at the centre of a galaxy, and then, due to some un-identified instability mechanism (see below for more discussions), they will transfer inward rapidly on to the black hole (BH) in a brief but luminous outburst \citep{Bailey1980, Shields1978}, leading to the so-called AGN phase of the duty cycle.
	
	Among the variability analysis, the shape of the long-term light curve can be served to diagnose the physical mechanism. For example, the light curve of a typical TDE follows a $t^{-5/3}$ power-law form \citep[][]{Rees1988, Phinney1989, Gezari2009}. Another example comes from gamma-ray bursts, which exhibit broken power laws with power-law index difference from one phase to the other, superimposed with various flares \citep{Zhang2006}. Interestingly, we notice that in X-ray binaries (XRBs), those dwarf novae and soft X-ray transients exhibit numerous outbursts, where each outburst shows a fast-rise-exponential-decay (FRED) light-curve profile in X-rays (e.g., \citealt{Chen1997,Powell2007,Yan2015c}). Physically, the FRED profiles in XRBs can be nicely interpreted under the thermal-viscous disc instability model (DIM; for reviews see \citealt{Lasota2001, Done2007}). To our knowledge, additional systems with the FRED profile are, some GRBs \citep{Peng2010}, the thermonuclear burst of one neutron star XRB \citep{Galloway2008}, the outburst of intermediate BH candidate ESO 243-49 HLX-1\citep{Lasota2011}, and two decade-long sustained TDE candidates \citep{Lin2017a,Lin2017}.
	
	In this work, we gather from literature the X-ray observations of the low-luminosity AGN (LLAGN) NGC 7213 over the past four decades. We report that this source also exhibits a FRED light curve, together with a possible weak reflare as caught by {\it RXTE}. Remarkably, the rising phase is also observed. NGC 7213 is the {\it first} normal AGN, to our knowledge, with the FRED evolution pattern observed (cf. \citealt{Strotjohann2016} for X-ray light curves of AGN with large amplitude).
	
	This paper is organized as follows. Section 2 presents the basic properties of LLAGN NGC 7213, together with its FRED flare profile. Section 3 provides the TDE interpretation, which we argue most likely. Other mechanisms, e.g., radiation-pressure instability and DIM, are discussed in Sections 4 and 5, where we argue that they are unlikely. The final section is devoted to a brief summary.
	
	\section{Long-term X-ray Flare of LLAGN NGC 7213}

In this section we will first provide the basic properties of NGC 7213, and then investigate the long-term light curve of NGC 7213 in various wavebands. We find that the unusual FRED pattern is mostly evident in X-rays, while there is no useful information/constrain in other wavebands (except the decline in radio).
		
	\subsection{Basic properties of NGC 7213} \label{sec:basic}
	
	NGC 7213 is a nearby face-on LINER (low-ionization nuclear emission-line region; \citealt{Phillips1979,Filippenko1984}) located at a luminosity distance $d_{\rm L}$=22.8 Mpc, a distance derived from a flat cosmology ($H_0=67.8$ km s$^{-1}$ Mpc$^{-1}$ and $\Omega_{\rm M}=0.308$ from \citealt{Planck-Collaboration2016}) with a corrected redshift $z_{\rm corr.3K}=0.005145$ to the reference frame defined by the 3K cosmic microwave background\footnote{\url{http://ned.ipac.caltech.edu}}. The central BH mass is $\mbh = 8^{+16}_{-6}\times 10^7\ \msun$ \citep{Schnorr-Muller2014}, which is derived from the $\mbh-\sigma$ relation in \citet{Gultekin2009}. Based on observations at epochs after 2000, the bolometric luminosity estimated from broad spectrum is $L_{\rm bol} \simeq 0.9-1.8 \times 10^{43}\ {\rm erg\,s^{-1}}\sim 1 \times 10^{-3}\ M_8^{-1}\ L_{\rm Edd}$ (for details see \citealt{Emmanoulopoulos2012,Schnorr-Muller2014}), where $M_8 \equiv\mbh/10^8\ \msun$ and the Eddington luminosity (for the hydrogen fraction of 0.7) is defined as $L_{\rm Edd}=1.47\times 10^{46}\  M_8\ {\rm erg\,s^{-1}}$. NGC 7213 is intermediate in radio, between radio-loud and radio-quiet, possibly, because of relatively large beaming effect, as the viewing angle in this system is likely small. The viewing angle of the jet can be constrained from that of the clumpy dusty torus, if we assume they are perpendicular to each one. Based on high-spatial-resolution infrared (IR) observations by Gemini, \citet{Ruschel-Dutra2014} modelled the IR spectrum and constrained the viewing angle to be $i\simeq 21\degr^{+9}_{-12}$.
	
	This source reveals a number of interesting properties. Observations after 2000 find that the ultraviolet bump is either absent or very weak in this source \citep{Starling2005,Lobban2010}. There is also no evidence for a Compton reflection continuum in hard X-rays, and the observed narrow Fe K$\alpha$ line is probably produced in the broad-line region \citep{Bianchi2003,Bianchi2008,Starling2005,Lobban2010, Ursini2015}. All these results suggest that a cold disc, if exists, should be truncated at a large radius after 2000 \citep{Starling2005,Lobban2010,Emmanoulopoulos2012}. We also find a complex (hybrid) correlation between the monochromatic radio luminosity $\lr$ and the 2--10 keV X-ray luminosity $\lx$, i.e. the correlation is unusually weak with $p\sim 0$ (in the form $\lr\propto\lx^p$) when $\lx$ is below a critical luminosity, and steep with $p>1$ when $\lx$ is above that luminosity \citep{Bell2011,Xie2016}. On the other hand, we also find, from its long-term X-ray spectral evolution, likely a V-shaped index--$\lx$ relation \citep{Xie2016}, i.e., its X-ray spectrum shows a `harder-when-brighter' behaviour when $\lx$ is low (first reported by {\it RXTE} observations; see \citealt{Emmanoulopoulos2012}), and an opposite behaviour when $\lx$ is high. The critical luminosity for the turnover in the $\lr$--$\lx$ correlation, estimated to be $L_{\rm X, crit} \approx 1.5\times 10^{42}\ \ergs$, is consistent with that constrained by the turnover in the index--$\lx$ correlation. Under the accretion--jet model \citep{Yuan2014}, which has been applied successfully to LLAGN and BH XRBs in their hard (and intermediate) states, \citet{Xie2016} then speculate that the accretion mode has been changed below and above this critical luminosity, i.e. it is a luminous hot accretion flow below $L_{\rm X, crit}$, and a two-phase (hot gas embedded with abundant cold clumps) accretion flow above it. Moreover, they also successfully modelled the broad-band (radio up to $\gamma$-rays) spectrum \citep{Xie2016}.

	\subsection{Long-term FRED light curve in X-rays}\label{sec:lc}
	\begin{table}
		\centering
		\caption{X-ray observations of NGC 7213}
		\label{fx}
		\begin{tabular}{ccccc} 
			\hline
			Time  & $F_\mathrm{X}$ (2--10 keV)&$\Gamma$ &Satellite  & Refs. \\
			(MJD)&{\tiny $10^{-11}$ erg/s/cm$^{2}$} & & &\\
			\hline
			43985  & 2.20  & 1.85$\pm$0.11  & $Einstein$  &  [1] \\
			44168  & 1.67  & 2.1$\pm$0.6      &  $Einstein$ &  [1]\\
			44374  &  6.80 & 1.72$\pm$0.12  &  $Einstein$ & [1]\\
			45644  &  5.00 & 1.78$\pm$0.05  & $EXOSAT$ & [2]\\
			48065  &  3.53$\pm$0.19 & 1.70$\pm$0.04  & $GINGA$    & [3]  \\
			48193  &  4.24$\pm$0.24 &  1.83$\pm$0.04 &  $GINGA$   & [3]  \\
			49479  &  3.01 &  1.73                  & $ASCA$   & [4]  \\
			52058  &  2.20   &  1.69$\pm$0.04 & $XMM$-$Newton$ & [5] \\
			       &          &               &   \& $BeppoSAX$   &    \\
			53846 & 2.71$\pm$0.02 & 1.80$\pm$0.02    &$RXTE$ & [6] \\
			53942 & 2.73$\pm$0.02 & 1.82$\pm$0.02    &$RXTE$ & [6] \\
			54030 & 2.44  & 1.75$\pm$0.02   & $Suzaku$ & [7] \\
			54037 & 2.50$\pm$0.02 & 1.84$\pm$0.02    &$RXTE$ & [6] \\
			54135 & 2.35$\pm$0.02 & 1.85$\pm$0.03    &$RXTE$ & [6] \\
			54234 & 2.31$\pm$0.02 & 1.84$\pm$0.03    &$RXTE$ & [6] \\
			54318  &  2.32$\pm$0.04 & 1.69$\pm$0.01 &  $Chandra$  & [8] \\
			54336 & 2.44$\pm$0.02 & 1.84$\pm$0.03    &$RXTE$ & [6] \\
			54430 & 2.16$\pm$0.02 & 1.87$\pm$0.03    &$RXTE$ & [6] \\
			54530 & 1.75$\pm$0.02 & 1.85$\pm$0.03    &$RXTE$ & [6] \\
			54635 & 1.18$\pm$0.02 & 1.90$\pm$0.05    &$RXTE$ & [6] \\
			54729 & 1.65$\pm$0.02 & 1.88$\pm$0.04    &$RXTE$ & [6] \\
			54821 & 1.86$\pm$0.02 & 1.87$\pm$0.03    &$RXTE$ & [6] \\
			54915 & 1.77$\pm$0.02 & 1.90$\pm$0.03    &$RXTE$ & [6] \\
			55102 & 1.54$\pm$0.02 & 1.90$\pm$0.04    &$RXTE$ & [6] \\
			55045 & 1.59$\pm$0.02 & 1.89$\pm$0.04    &$RXTE$ & [6] \\
			55039 & 1.33$\pm$0.02 & 1.91$\pm$0.04    &$RXTE$ & [6] \\
			55147 & 1.23$\pm$0.01 & 1.86$\pm$0.02    & $XMM$-$Newton$  & [9] \\
			56935 & 1.60$\pm$0.10 & 1.84$\pm$0.03    & $NuSTAR$ &[10]\\
			55493 & 1.05$\pm$0.14 & 1.67$\pm$0.11    & $Swift$ &[11]\\
			55751 & 1.32$\pm$0.28 & 1.68$\pm$0.17    & $Swift$ &[11]\\
			55756 & 0.99$\pm$0.19 & 1.87$\pm$0.15    & $Swift$ &[11]\\
			56937 & 1.42$\pm$0.18 & 1.66$\pm$0.10    & $Swift$ &[11]\\
			57752 & 1.18$\pm$0.19 & 1.62$\pm$0.13    & $Swift$ &[11]\\
			\hline
		\end{tabular}
		References: [1]\citet{Halpern1984}; [2]\citet{Turner1989}; [3]\citet{Nandra1994}; [4]\citet{Turner2001}; [5]\citet{Bianchi2003}; [6]\citet{Emmanoulopoulos2012}; [7]\citet{Lobban2010}; [8]\citet{Bianchi2008}; [9]\citet{Emmanoulopoulos2013}; [10]\citet{Ursini2015}; [11]this work
	\end{table}

	\begin{figure*}
		\centering
		\includegraphics[width=0.7\linewidth]{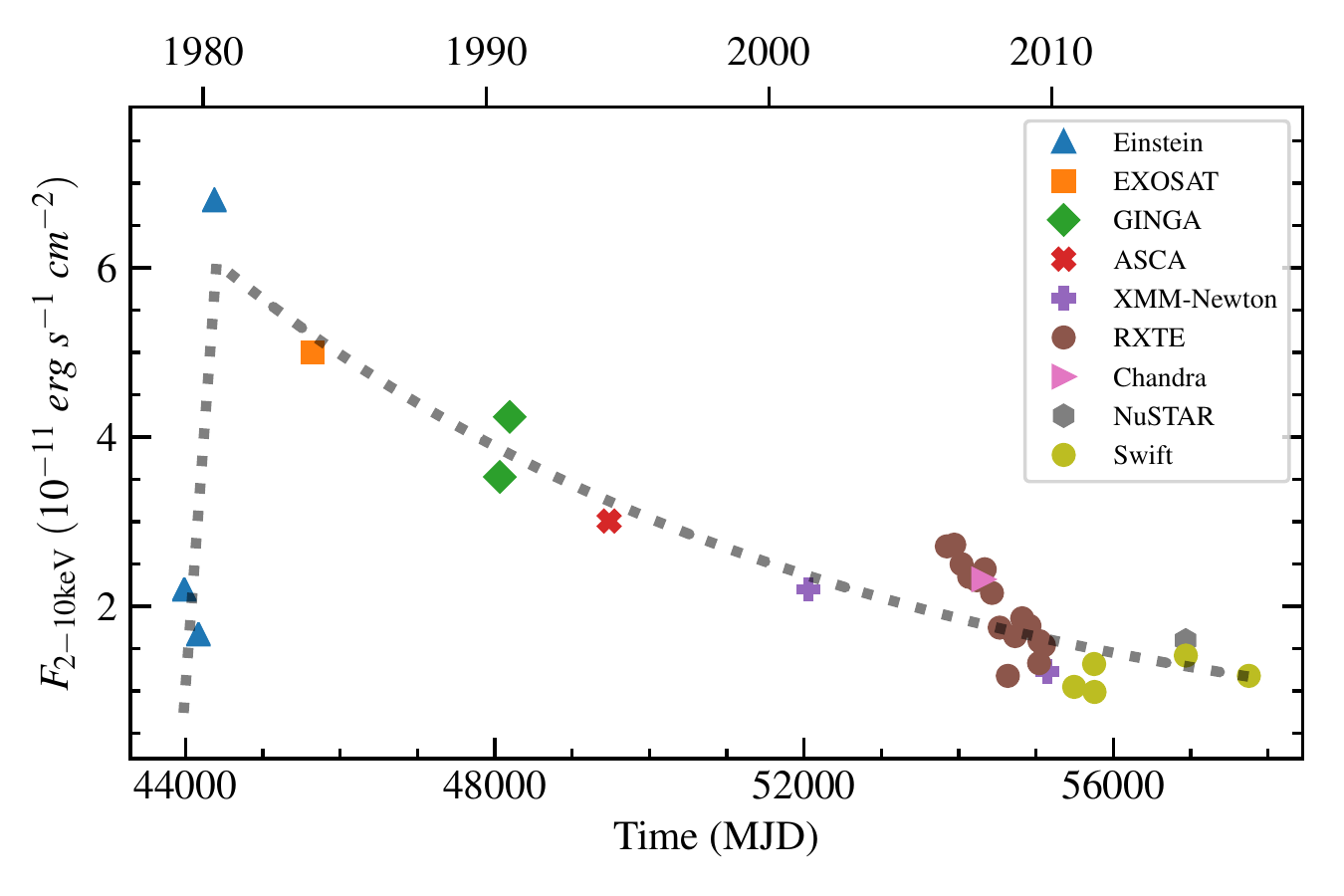}
		\caption{The long-term X-ray light curve in LLAGN NGC 7213. This source brightens by a factor of $\sim 4$ within $\sim$ 200d (from MJD 44168 to MJD 44374) and then declines exponentially afterwards, with a possible weak flare, as detected by {\it RXTE}. For clarity, uncertainties in the observational data are not shown here. The dotted line shows the FRED profile fitting to the observational data, where the $e$-folding decay time-scale is constrained to be $\tau_{\rm decay}\approx8116$d ($\approx22.2$ yr). As labelled in the figure, different symbols mark observations from different X-ray satellites.}
		\label{lc}
	\end{figure*}

	\autoref{lc} shows the long-term light curve of NGC 7213 in hard X-rays over the past four decades, from modified Julian date (MJD) 43985 to MJD 57752. We list in \autoref{fx} the observation time (MJD), the 2-10 keV X-ray flux $F_{\rm X}$, the X-ray photon index $\Gamma$ (defined as $F_{\rm E}\propto E^{1-\Gamma}$, where $E$ is the photon energy), as well as the corresponding satellites of the observations. Most of the data are collected from \citet{Bell2011,Ursini2015} and references therein.

	In X-rays, NGC 7213 was first identified by $HEAO$-$1$ in 1977 \citep{Marshall1979}. However, due to the lack of the spectral information \citep{Piccinotti1982}, we exclude them from our sample (cf. \autoref{fx}). Since then, NGC 7213 has been observed by almost every X-ray satellite. In this work, we are mainly interested in its long-term evolution, thus we focus on the 2--10 keV band, which is covered by most X-ray missions. X-ray observations, which do not cover this energy band, e.g., those by $ROSAT$ and $CGRO$, are thus not included in \autoref{fx}. The continuum spectrum in 2--10 keV is all well fitted by a power-law component, while photon index varies in the range $\sim$ 1.6--2.1 (see \autoref{fx}). We take the rebinned data set for the $RXTE$ observations, with each bin $\sim$100d long (see details in \citealt{Emmanoulopoulos2012}). In this case, the MJD reported in \autoref{fx} is the average value of each bin for the $RXTE$ data.
	
	NGC 7213 is also observed frequently by the $Swift$/XRT \citep{Burrows2005}, which has never been reported. We hereby briefly describe the $Swift$/XRT data analysis. The $Swift$/XRT event data were first processed with XRTPIPELINE (v0.13.2) to generate the cleaned event data. Then we extracted the spectra by using XSELECT from a region of a circle when observed in the photon counting (PC) mode and a box when observed in the windowed timing (WT) mode. The $Swift$/XRT spectra were all well fitted with an absorbed power-law model by using XSPEC 12.9.0. To be consistent with data from the literature, the hydrogen column density $N_{\rm H}$ was fixed to 2.04$\times 10^{20}$ cm$^{-2}$ \citep{Ursini2015}. The 2--10 keV X-ray flux can then be derived from the spectral fitting results, as reported in \autoref{fx}.

	From \autoref{lc}, we find that the X-ray flux increased by a factor of $\sim 4$ in less than one year ($\sim$ 200d, between MJD 44168 and MJD 44374, all are observed by {\it Einstein} satellite, cf. \autoref{fx}), and then decreased gradually in the following 30 years. The onset/trigger of the flare may be missed, as the flux before the peak is higher than the lowest fluxes during the decay phase (by 2016 December 30), when obviously the source still does not reach its quiescence. The peak X-ray luminosity of this flare is $L_{\rm X, peak}\approx 4.2\times 10^{42}\ \ergs$ (equivalently, $L_{\rm X, peak}\approx 2.9\times 10^{-4}\ M_8^{-1}\ \ledd$). Correspondingly, the peak bolometric luminosity can be estimated as $L_{\rm bol, peak} \approx 6.8\times10^{43} (f_{\rm X}/16)\ \ergs$, or equivalently $L_{\rm bol, peak}\approx 4.6\times 10^{-3} (f_{\rm X}/16)\ M_8^{-1}\ \ledd$, which is about a factor of $5$ brighter than estimations based on later observations \citep[e.g.][]{Emmanoulopoulos2012,Schnorr-Muller2014}. Here, the X-ray bolometric correction $f_{\rm X}$ is set to $f_{\rm X}\equiv L_{\rm bol}/\lx\approx 16$, an `average' value suggested by observations of LLAGNs \citep{Ho2008}.
	
	As shown in Fig. \ref{lc}, the profile of X-ray light curve is quite similar to a FRED form, where both the rise and the decay phases are observed. To quantitatively examine this, we adopt the following FRED function (\autoref{fred}) to model the X-ray light curve,
	\begin{equation}
	\label{fred}		
	F_{\rm X}(t) = \left\{ \begin{array}{ll} \frac{t-t_\mathrm{s}}{t_\mathrm{p}-t_\mathrm{s}}\, F_{\rm X, peak} & (t\leq t_\mathrm{p}) \\
			e^{-\frac{t-t_\mathrm{p}}{\tau_{\rm decay}}}\, F_{\rm X, peak} & (t\geq t_\mathrm{p})\\
			\end{array}  \right.
	\end{equation}
	where the $t_\mathrm{s}$ and $t_\mathrm{p}$ are the start and peak time of the flare, $F_{\rm X, peak}$ and $\tau_{\rm decay}$ are the peak flux and decay time-scale. As shown by the dotted line in \autoref{lc}, the light curve agrees with FRED fairly well. The $e$-folding decay time-scale is constrained to be $\tau_{\rm decay} \approx8116$d ($\approx 22.2$ yr). We note that {\it RXTE} may possibly detect a weak reflare that lasts $\sim$3yr (between 2006 July and 2009 September; \citealt{Emmanoulopoulos2012}).

\subsection{Long-term evolution information from other wavebands}
\label{sec:oir}

NGC 7213 is a bright galaxy in the optical/infrared band, with numerous photometry observations in the literature from as early as 1950s \citep[e.g.][]{Evans1952}. However, the aperture size in those early works is too large to obtain the nuclear contribution. For example, the bulge component dominates the emission at the $J$ band even within 1$\arcsec$ \citep{Prieto2002}. Broad lines in optical band are indeed observed (e.g., \citealt{Phillips1979,Filippenko1984}), but they can mainly provide constrains in BH mass. Practically, there is no useful information on the long-term evolution of NGC 7213 in the optical/infrared band.

NGC 7213 was first identified in the radio band by the low-resolution single-dish {\it Parkes} 2700 MHz Survey \citep{Wright1974,Wright1977}. These detections are before the rise of the X-ray flux. The 2.7 GHz radio flux observed by {\it Parkes} is roughly stable \citep{Wright1977,Halpern1984,Sadler1984}. We caution that there exists a circumnuclear ring of H {\sc ii} regions at $\sim 2$ kpc from the nucleus \citep{Storchi-Bergmann1996}, and the H {\sc ii} regions are known to be radio emitters. We did not find any radio observations that cover the period of the fast-rise and the early decay phases of the X-ray emission ($\sim$ MJD 44000--47000). Higher frequency radio observations at 8.4 GHz since 1988, much later after the peak in X-rays, do find an obvious decline trend \citep{Blank2005,Bell2011}. There is also a weak reflare detected in the radio band, which is simultaneous with the X-ray reflare detected by $RXTE$ \citep[2006--2009;][]{Bell2011}. The long-term radio flux in the 8.4 GHz positively correlates with X-rays in a complex way (cf. \autoref{sec:basic}; \citealt{Bell2011, Xie2016}).

	\section{A Delayed TDE of A Main-Sequence Star} \label{sec:tde}
	We in this section investigate the TDE scenario for the nuclei activity in NGC 7213. When a star approaches the SMBH at a distance comparable to the tidal radius ($r_t = R_{*} (M_{\rm BH}/M_{*})^{1/3}$; cf. \citealt{Ulmer1999}), immense tidal forces from the SMBH will (partially) disrupt it, resulting in a stream of debris that falls back on to the SMBH and powers a luminous flare. The maximal mass fall-back rate decreases with the BH mass \citep{Evans1989,Ulmer1999,Guillochon2013}:
	\begin{equation}
		\dot{M}_\mathrm{fb,peak} \propto M_\mathrm{BH}^{-1/2}M_{*}^{2}R_{*}^{-3/2}
	\end{equation}
	where $M_{*}$ and $R_{*}$ are, respectively, the mass and radius of the star. Obviously, with $\dot{M}_\mathrm{fb,peak} c^2/L_{\rm Edd}\propto M_\mathrm{BH}^{-3/2}$, it can be sub-Eddington when $M_{\rm BH}\sim10^8\ \msun$, the case in NGC 7213. This makes the TDE scenario attractive. For example, the lack of prominent thermal emission, a component observed in most TDEs, is reasonable in this source, as it may stay in hot accretion flow mode due to low accretion rates \citep{Yuan2014}. Another advantage of TDE scenario is that the resulting accretion disc of a TDE is very compact, i.e. $R \la 500\, R_{\rm s}$ (here $R_{s}=2GM_{\rm BH}/c^{2} = 3\times10^{13}\, (M_{\rm BH}/10^{8})$ cm is the Schwarzschild radius of BH), compared to that of normal accretion discs around SMBHs. Consequently, in X-rays we should not observe any reflection emission from cold accretion gases, which is in excellent agreement with X-ray observations (\citealt{Bianchi2003,Bianchi2008,Starling2005,Lobban2010, Ursini2015}, cf. \autoref{sec:basic}). Besides, the ionizing gas inferred from narrow Fe K$\alpha$ line in X-rays may just be the unbound debris of the TDE. We additionally note that the observed broad Balmer lines (e.g., H$\alpha$ and H$\beta$ series) shortly after the flare in this system \citep{Phillips1979,Filippenko1984,Schnorr-Muller2014} are consistent with the TDE interpretation. This is mainly because the broad line region is close to the BH, at a distance within $\sim 1-10$ light days (or typically hundreds to thousands of $R_{\rm s}$; \citealt{Peterson2006}). Despite the uncertainty in the formation of broad line clouds, there is sufficient time either for the wind/outflow launched from the central accretion disc to dynamically move to that distance, or to light and accelerate the pre-existing clouds there.
	
	However, there are still two challenges for the TDE interpretation: one is the duration of the flare and the other is the FRED profile of light curve in X-rays. The duration of most TDEs observed is usually short, of the order of months to one year (for a review on TDE observations, see \citealt{Komossa2015a}), which is much shorter than the 40yr duration observed in NGC 7213. However, long-duration TDE scenario has been applied to understand the long-term X-ray flares in several sources, i.e. IC 3599 \citep{Campana2015}, NGC 3599 \citep{Saxton2015}, 3XMM J150052.0+015452 \citep{Lin2017a} and 2XMM J123103.2+110648 \citep{Lin2017}, where the whole duration of the latter two is over one decade. Interestingly, unlike those normal TDEs whose decay profiles are power law (e.g., $t^{-5/3}$; \citealt{Rees1988,Gezari2009}), the decay light curve profile of the two decade-long TDEs \citep{Lin2017a,Lin2017} is also exponential, with the decay time-scale $\tau_{\rm decay}$ more than 1000d.
	
	To have a TDE that can last for years to decades, additional mechanism such as slow circularization should operate, i.e. instead of an efficient orbital circularization, the fall-back stream of disrupted gas gradually circularizes to form an accretion disc around the central BH under the viscous influence, after which radiation will emit out (e.g., \citealt{Guillochon2014,Guillochon2015, Shiokawa2015,Hayasaki2016}). The accretion will then be determined by the viscous time-scale $\tau_\mathrm{visc}$, which is much longer than the free fall-back time-scale $\tau_\mathrm{fb}$. Interestingly, with the delay between the stream falling back and the final accretion on to BH, both the long decay time-scale and the FRED light curve can be realized simultaneously (see the supplementary note 7 in \citealt{Lin2017a}). Due to the pile-up of accreting gas, the delayed mass accretion rate $\dot{M}_\mathrm{acc}$ naturally shows an exponential decay profile. Following \citet{Lin2017a}, the delayed mass accretion rate $\dot{M}_\mathrm{acc}$ can be expressed as,
	 \begin{equation}
	 \label{M_acc}
	 	\dot{M}_\mathrm{acc}(t) = \frac{1}{\tau_\mathrm{visc}}\Big( e^{-t/\tau_\mathrm{visc}}\int^{t}_{0}e^{t'/\tau_\mathrm{visc}}\dot{M}_\mathrm{fb}(t')dt'\Big)
	 \end{equation}
	 The evolution of the mass fall-back rate $\dot{M}_\mathrm{fb}$ in \autoref{M_acc} can be determined through hydrodynamical simulations of the tidal disruption process \citep[e.g.][]{MacLeod2012,Guillochon2013,Guillochon2014}. The viscous time-scale $\tau_\mathrm{visc}$ can be expressed as \citep{Guillochon2013}
	 \begin{equation}
	 \label{t_visc}
	 	\tau_\mathrm{visc} = 3.2\times10^{-3} \beta^{-3}\Big(\frac{\alpha}{0.1}\Big)^{-1}\Big(\frac{M_{*}}{M_{\odot}}\Big)^{-1/2}\Big(\frac{R_{*}}{R_{\odot}}\Big)^{3/2}~{\rm yr}
	 \end{equation}
	 where $\alpha$ is the viscosity parameter and $\beta$ is the ratio between the tidal radius $r_\mathrm{t}$ and the pericentre distance $r_\mathrm{p}$, $\beta=r_\mathrm{t}/r_\mathrm{p}$.
	
	\begin{figure*}
		\centering
		\includegraphics[width=0.7\linewidth]{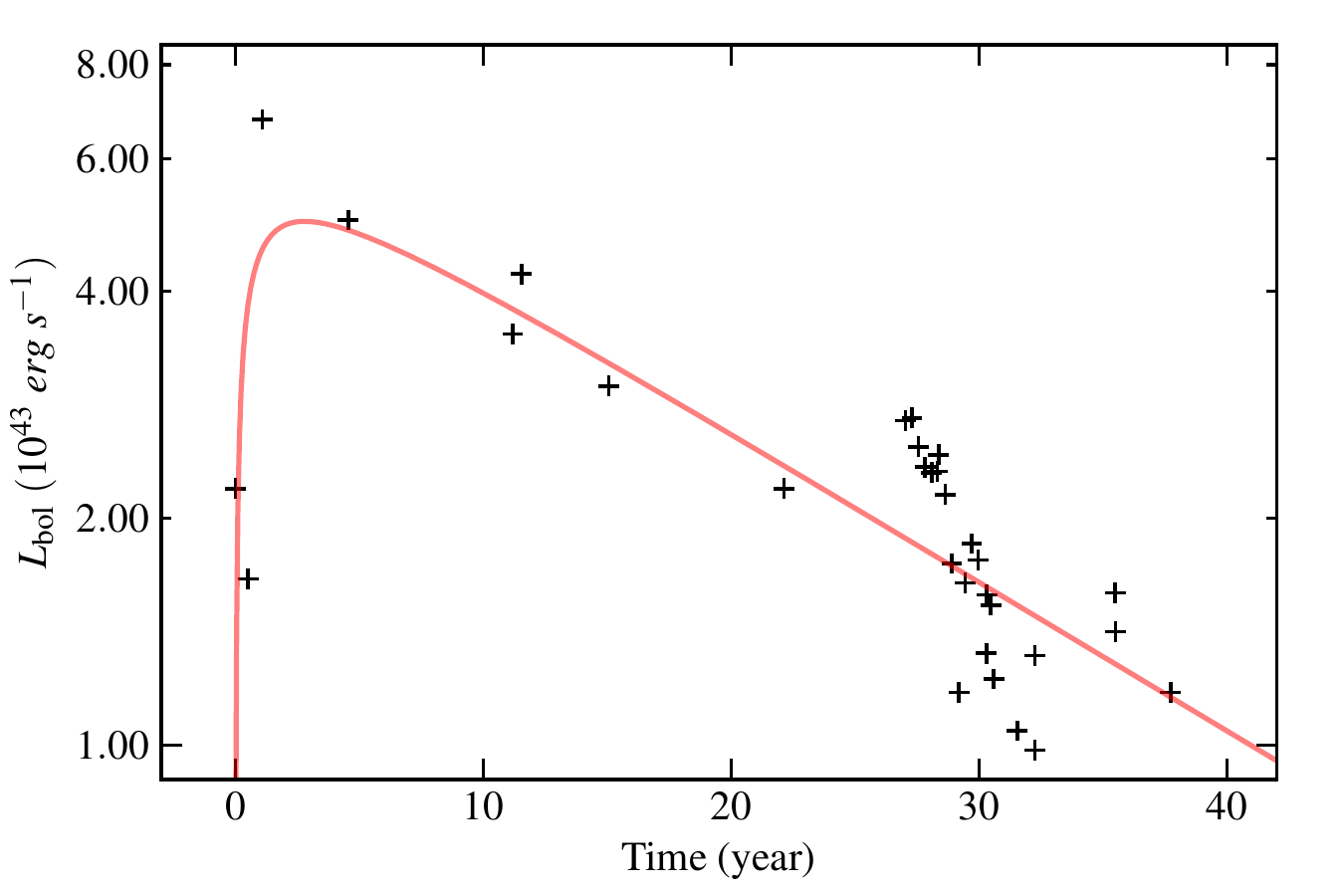}
		\caption{Theoretical modelling of the light curve based on the delayed TDE scenario. We assume the X-ray bolometric correction to be $f_{\rm X}=16$ (cf. \autoref{sec:lc}) and the radiative efficiency to be $\epsilon=0.1$. The red line represents the modelling of a delayed TDE, with parameters $M_\mathrm{BH}=10^{8}\ M_{\odot}$ (fixed), $\beta=1.0$ (fixed), $M_{*}=4.27\ M_{\odot}$ (the corresponding $R_{*}=2.41R_{\odot}$ for a main-sequence star) and $\alpha=2.85\times10^{-5}$. See text for details. }
		\label{fit}
	\end{figure*}
	
	Below we investigate whether a delayed TDE can reproduce the FRED light curve observed in NGC 7213. We assume the X-ray bolometric correction to be $f_{\rm X}=16$ and the radiative efficiency to be $\epsilon=0.1$ cf. \autoref{sec:lc}, in order to compare the modelling light curves with observations. We simplify the modelling by using the analytical approximations in \citet{Guillochon2013,Guillochon2014}, i.e. we take a power-law dependence of the mass fall-back rate $\dot{M}_\mathrm{fb}$ on $M_\mathrm{BH}$, stellar mass $M_{*}$ and stellar radius $R_{*}$ as \citep{Guillochon2014}
\begin{equation}
\dot{M}_{fb} = \Big(\frac{M_\mathrm{BH}}{10^{6}\ M_{\odot}}\Big)^{-1/2}\Big(\frac{M_{*}}{M_{\odot}}\Big)^{2}\Big(\frac{R_{*}}{R_{\odot}}\Big)^{-3/2}\dot{M}_\mathrm{init}(\beta).
\end{equation}
Here $\dot{M}_\mathrm{init}(\beta)$ is the mass fall-back rate of a solar type stellar disrupted by a $10^{6}M_{\odot}$ SMBH, which comes from detailed numerical simulations in \citet{Guillochon2013}. In order to derive the $\dot{M}_\mathrm{fb}$, we also take the interpolation method for different impact parameter $\beta$ \citep{Guillochon2014}. We assume the disrupted star is a main-sequence star, with mass-radius relation given in \citet{Tout1996}, and the polytropic index of the disrupted star to be $\gamma=5/3$. With fitting formulae given above, the impact of different parameters can be understood easily. For example, with other parameters fixed, smaller $\beta$ ($\lesssim1$) means the pericentre of the star moves outwards. Consequently, $\tau_\mathrm{visc}$ will be larger, i.e. we will have a longer duration flare. Meanwhile, the tidal force will be weaker, thus star could eventually survive the encounter and a smaller fraction of its mass will be captured by the BH \citep{Guillochon2013}. The dependences on $\beta$ for the evolutions of many quantities have been investigated in \citet{Guillochon2013}. Here we simply chose a typical value $\beta = 1.0$. The BH mass is fixed as $M_\mathrm{BH} = 10^{8} M_{\odot}$ \citep{Blank2005,Schnorr-Muller2014}. We adopt the `Emcee' implementation \citep{Foreman-Mackey2013} of the Markov Chain Monte Carlo (MCMC) sampler to constrain the other two parameters $M_{*}$ and $\alpha$. From the posterior samples, we obtain that $M_{*}=4.27\pm1.55M_{\odot}$ (the corresponding $R_{*}=2.41\pm0.49R_{\odot}$) and an extremely low viscous parameter $\alpha=2.85\pm1.44 \times10^{-5}$. The corresponding viscous time-scale is about 20.34 yr (\autoref{t_visc}), which is similar to the observed $e$-folding decay time-scale. The light curve of a delayed TDE with the best-fitting parameters ($M_{*}=4.27M_{\odot}$ and $\alpha=2.85\times10^{-5}$) is shown as the red solid curve in \autoref{fit}, which is consistent with the observed light curve. We admit that the weak reflare during the decay is not taken into account during the modelling. To conclude, we find that a delayed TDE of a normal star is viable to produce the FRED light curve in X-rays as observed in NGC 7213.
	
	At last, we need to discuss the AGN activity of NGC 7213. We obviously cannot rule out the possibility of the pre-TDE nuclear activity, especially considering the detections in X-ray and radio band before the onset of the flare \citep{Marshall1979,Piccinotti1982,Wright1974,Wright1977}. The expected TDE rate in AGN is much higher than the quiescent galaxy \citep{Karas2007,Kennedy2016}. However, only a few flares/outbursts in AGN have been considered as TDEs. For example, the long-duration flares in the two AGN, NGC 3599 and IC 3599, have been argued as TDEs \citep{Saxton2015,Campana2015}. The host galaxy of typical TDE ASASSN-14li has also been detected in radio band long before the outburst, which indicates the possible AGN activity \citep{Alexander2016}. The coincidence of the optical transient CSS100217:102913+404220 with the nuclear of its host galaxy makes it as a candidate TDE in the narrow-line Seyfert 1 galaxy (NLS1). Recently, the transient PS16dtm was interpreted as a TDE in the NLS1 galaxy \citep{Blanchard2017a}. The newly formed accretion disc during the TDE may change the geometry of the pre-existing accretion disc, which causes an unusual evolution of X-ray luminosity \citep{Blanchard2017a}. There are some studies that are trying to investigate the effects brought by the presence of the pre-existing accretion flow. \citet{Bonnerot2016} found that the ambient gas on the TDE bound debris stream, in some cases, will dissolve a significant part of the stream. Interestingly, the interactions between the debris stream and the pre-existing accretion flow can potentially stall the stream far from the SMBH, and lead to a dim and long flare \citep{Kathirgamaraju2017}. For example, the accretion time-scale will be $\sim$10 yr and the corresponding accretion rate will be 0.01 $L_\mathrm{Edd}$ for a $10^{6}$ SMBH  \citep{Kathirgamaraju2017}.

	\section{A newly triggered AGN}
	Based on the radio spectral properties, it is argued that NGC 7213 is a gigahertz-peaked spectrum (GPS) source \citep{Blank2005, Hancock2009}. The compact radio emission site ($<3$ mas, or equivalently $< 0.33$ pc; cf. \citealt{Blank2005}) implies that NGC 7213 could possibly be a newly triggered young radio galaxy.
	
	One scenario for the trigger of the short-lived GPS sources is the radiation-pressure-driven instability proposed by \citet{Czerny2009}. This model has been applied to numerous systems, e.g., the repetitive flares with time-scales of tens of seconds observed in two BH XRBs, GRS 1915$+$105 and IGR J17091$-$3624 \citep{Belloni2000,Altamirano2011}, and one NS XRB, MXB 1730$-$335 \citep{Bagnoli2015}, the short-lived compact young radio sources \citep{Czerny2009, Wu2009}, the flares in AGNs IC 3599 and NGC 3599 \citep{Grupe2015,Saxton2015} and even the outbursts in ultra-luminous sources ESO 243$-$49 HLX-1 \citep{Sun2016,Wu2016}. We caution that the flares in the two AGNs (IC 3599 and NGC 3599) are also inferred to be TDEs \citep{Montesinos-Armijo2011,Saxton2015,Campana2015}.
	
	However, from theoretical point of view, the presence of this instability is actually under active debate, e.g., opposite results can be found in different numerical simulations \citep{Hirose2009,Jiang2013,Mishra2016,Sc-adowski2016}. Apart from this debate, one prediction of the radiation-pressure-driven instability model is that there exists a universal correlation among the duration of the outburst, the peak bolometric luminosity, and the viscosity parameter of the accretion disc $\alpha$ \citep{Czerny2009,Wu2016}, i.e., $\log (T_{\rm burst}/{\rm yr}) = 1.25\log (L_{\rm bol, peak}/\ergs) + 0.38\log(\alpha/0.02)-53.6$. Interestingly, the flare observed in NGC 7213, with a duration of $\sim$20 yr, peak bolometric luminosity $\sim 7\times 10^{43}\ergs$ (see \autoref{sec:basic}), and $\alpha\approx 0.01$ \citep{Xie2016}, agrees with this correlation.
	
  However, there are several pieces of evidence against the radiation-pressure instability model. First, the radiation-pressure instability should occur in the high-luminosity regime \citep{Shakura1973,Lightman1974,Janiuk2002}. For example, the average luminosity of the sample in \citet{Wu2016} is about 0.3 $L_\mathrm{Edd}$. \citet{Czerny2009} gave a threshold accretion rate for the radiation-pressure instability $\sim0.025 \dot{M}_\mathrm{Edd}$.  However, the peak luminosity of this flare of NGC 7213 is still lower than the luminosity required by the radiation-pressure instability. Secondly, a slow-rise-fast-decay light curve is the characteristic property of the radiation-pressure instability model \citep{Janiuk2002, Janiuk2004, Janiuk2011, Czerny2009, Grzedzielski2017}. For example, the duration for which the flux changes by one order of magnitude during the rise phase is two or three times longer than that during the decay phase (see the simulated light curves for a $10^8~M_{\odot}$ SMBH in fig. 3 of \citealt{Czerny2009}). This is contradictory to that observed in NGC 7213, where (decay time-scale)/(rise time-scale) $>10$ (cf. \autoref{lc}). Thirdly, the typical amplitude of an SMBH flare driven by radiation-pressure instability is about 2-3 orders of magnitude \citep[e.g.][]{Czerny2009}; for comparison, the amplitude in NGC 7213 is only $\sim $5-6 (cf. \autoref{sec:lc}). Smaller amplitude flare can be produced on the condition that the viscous stress of angular momentum transport has a stronger dependence on the gas pressure than the total (gas+radiation) pressure \citep{Grzedzielski2017}, However, such dependence seems unlikely, as the magneto-hydrodynamical simulations confirm that the viscous stress should be proportional to the total pressure, not other variants \citep{Hirose2009,Blaes2014}. Considering the above reasons, we disfavour the operation of radiation-pressure instability in NGC 7213.

	\section{Thermal-viscous disc instability model}\label{sec:dim}
	The thermal-viscous DIM model (see review in \citealt{Lasota2001}) is first to explain the outburst of dwarf nova and then is applied to low-mass X-ray transients (LMXBTs). Although the idea of applying this disc instability in AGN was proposed decades ago (e.g., \citep{Lin1986, Siemiginowska1996, Burderi1998, Hameury2009}), there is still no evidence of this mechanism to operate in any AGN. The DIM can naturally produce a FRED light curve \citep{Cannizzo1994,Cannizzo1995,King1998}. The extended exponential decay will be achieved when the disc can stay in hot state and maintains a quasi-steady surface density profile during the outburst \citep{King1998,Lasota2016}. The decay time-scale ($\tau_\mathrm{decay}$) is determined by the viscous time-scale at the critical radius for instability, i.e. $\tau_\mathrm{decay} \approx \frac{1}{3}\tau_\mathrm{visc}$ \citep{King1998}. The critical radius is given as \citep[e.g. Eq.2 in][]{Lasota2012}:
		\begin{equation}
			R_\mathrm{crit}\approx 42R_\mathrm{s}\Big(\frac{\dot{M}}{10^{-2}\dot{M}_\mathrm{Edd}}\Big)^{1/3}\Big(\frac{M}{10^{8}M_{\odot}}\Big)^{-1/3}
		\end{equation}
		where $\dot{M}_\mathrm{Edd} = 10L_\mathrm{Edd}/c^{2}$ is the Eddington accretion rate. Numerical calculation also shows that the critical radius for instability is orders of magnitude of 100 $R_\mathrm{s}$ for a $10^{8}M_{\odot}$ BH at 0.01 $\dot{M}_\mathrm{Edd}$ accretion rate \citep{Janiuk2004,Janiuk2011}. The viscous time-scale at the critical radius is \citep[e.g. Eq.6 in][]{Lasota2012}
		\begin{equation}
			\tau_\mathrm{visc} \sim 10^{6}\Big(\frac{\alpha}{0.1}\Big)^{-1}\Big(\frac{T}{10^{4}\mathrm{K}}\Big)^{-1}\Big(\frac{R}{10^{15}\mathrm{cm}}\Big)^{1/2}\Big(\frac{M}{10^{8}\mathrm{M}_{\odot}}\Big)^{1/2}~\mathrm{yr}
		\end{equation}
	Therefore, here is one serious problem with the DIM application in NGC 7213 that the decay time-scale predicted by DIM is much longer than that observed in NGC 7213. This is actually a well-known challenge for the application of DIM to AGNs \citep{Lin1986}. We thus disfavour the DIM model as the trigger of the flare observed in NGC 7213.

\section{Summary}
	Among the variability analysis, the shape of the long-term light curve can be served to diagnose its physical mechanism. We summarize the main observational characteristics of this X-ray flare of NGC 7213 as follows: This flare lasts nearly 40 years. The peak bolometric luminosity of this flare is $\sim$ 0.01 $L_\mathrm{Edd}$. The X-ray light curve shows a FRED profile (see \autoref{lc}). The $e$-folding decay time-scale is approximately $8116$d ($\approx 22.2$ yr). The amplitude of the flare is constrained to be larger than $\sim$ 5-6, as the trigger of this flare may still be missed and it has not recovered back to its quiescence (2016 Dec).
	
	We then examined the possible variability models proposed in the literature, and find that either the newly triggered AGN scenario (driven by radiation-pressure instability) or the thermal-viscous DIM fails to explain some key properties observed in NGC 7213, thus we ague them unlikely. For examples, the radiation-pressure instability is incapable to produce the FRED profile, and decay time-scale from the thermal-viscous DIM is several orders of magnitude longer than that observed in NGC 7213. A delayed TDE of a main-sequence star is favoured. The disruption of a star by a massive SMBH ($\sim10^{8}M_{\odot}$) can produce a long-duration flare with low peak luminosity, and a small fraction of mass is accreted by the SMBH in the case of small impact parameter \citep{Guillochon2013}. Additionally, if the disrupted bound stream suffers slow circularization, an exponential decay profile can be produced \citep{Lin2017a}. So the X-ray light curve of NGC 7213 fits the delayed TDE scenario quite well, as confirmed by our detailed light-curve modelling (\autoref{fit}). Under the TDE interpretation, the flare in NGC 7213 has several unique properties compared to others. First, TDEs normally happen around less massive SMBHs, while in NGC 7213 it is a partial disruption of a star around a $\mbh\sim 10^8\ \msun$ BH. Secondly, normally the peak bolometric luminosity of TDEs is around or above the Eddington luminosity, but this one is highly sub-Eddington, $L_{\rm bol, peak} \sim$ 0.01 $L_{Edd}$. Thirdly, the flare in NGC 7213 lasts more than four decades, which is much longer compared to others.

	\section*{Acknowledgments}
	
	We appreciate Paul Wiita, James Guillochon, Tinggui Wang, Wenfei Yu, Bin Liu, Morgan MacLeod and Dacheng Lin for helpful discussions, the referee for suggestions, and Dimitris Emmanoulopoulos for providing us with the {\it RXTE} data of NGC 7213. This work was supported in part by the National Key Research and Development Program of China (2016YFA0400804), the Youth Innovation Promotion Association of CAS (id. 2016243), the Natural Science Foundation of Shanghai (No. 17ZR1435800), and the National Natural Science Foundation of China (grant Nos. 11403074, 11773055, 11333005 and 11350110498). This research has made use of data and/or software provided by the High Energy Astrophysics Science Archive Research Center (HEASARC), which is a service of the Astrophysics Science Division at NASA/GSFC and the High Energy Astrophysics Division of the Smithsonian Astrophysical Observatory.
	

\begin{thebibliography}{}
\makeatletter
\relax
\def\mn@urlcharsother{\let\do\@makeother \do\$\do\&\do\#\do\^\do\_\do\%\do\~}
\def\mn@doi{\begingroup\mn@urlcharsother \@ifnextchar [ {\mn@doi@}
  {\mn@doi@[]}}
\def\mn@doi@[#1]#2{\def\@tempa{#1}\ifx\@tempa\@empty \href
  {http://dx.doi.org/#2} {doi:#2}\else \href {http://dx.doi.org/#2} {#1}\fi
  \endgroup}
\def\mn@eprint#1#2{\mn@eprint@#1:#2::\@nil}
\def\mn@eprint@arXiv#1{\href {http://arxiv.org/abs/#1} {{\tt arXiv:#1}}}
\def\mn@eprint@dblp#1{\href {http://dblp.uni-trier.de/rec/bibtex/#1.xml}
  {dblp:#1}}
\def\mn@eprint@#1:#2:#3:#4\@nil{\def\@tempa {#1}\def\@tempb {#2}\def\@tempc
  {#3}\ifx \@tempc \@empty \let \@tempc \@tempb \let \@tempb \@tempa \fi \ifx
  \@tempb \@empty \def\@tempb {arXiv}\fi \@ifundefined
  {mn@eprint@\@tempb}{\@tempb:\@tempc}{\expandafter \expandafter \csname
  mn@eprint@\@tempb\endcsname \expandafter{\@tempc}}}

\bibitem[\protect\citeauthoryear{{Alexander}, {Berger}, {Guillochon},
  {Zauderer}  \& {Williams}}{{Alexander} et~al.}{2016}]{Alexander2016}
{Alexander} K.~D.,  {Berger} E.,  {Guillochon} J.,  {Zauderer} B.~A.,
  {Williams} P.~K.~G.,  2016, \mn@doi [\apjl] {10.3847/2041-8205/819/2/L25},
  \href {http://adsabs.harvard.edu/abs/2016ApJ...819L..25A} {819, L25}

\bibitem[\protect\citeauthoryear{{Altamirano} et~al.,}{{Altamirano}
  et~al.}{2011}]{Altamirano2011}
{Altamirano} D.,  et~al., 2011, \mn@doi [\apjl] {10.1088/2041-8205/742/2/L17},
  \href {http://adsabs.harvard.edu/abs/2011ApJ...742L..17A} {742, L17}

\bibitem[\protect\citeauthoryear{{Ar{\'e}valo} \& {Uttley}}{{Ar{\'e}valo} \&
  {Uttley}}{2006}]{Arevalo2006}
{Ar{\'e}valo} P.,  {Uttley} P.,  2006, \mn@doi [\mnras]
  {10.1111/j.1365-2966.2006.09989.x}, \href
  {http://adsabs.harvard.edu/abs/2006MNRAS.367..801A} {367, 801}

\bibitem[\protect\citeauthoryear{{Bagnoli} \& {in't Zand}}{{Bagnoli} \& {in't
  Zand}}{2015}]{Bagnoli2015}
{Bagnoli} T.,  {in't Zand} J.~J.~M.,  2015, \mn@doi [\mnras]
  {10.1093/mnrasl/slv045}, \href
  {http://adsabs.harvard.edu/abs/2015MNRAS.450L..52B} {450, L52}

\bibitem[\protect\citeauthoryear{{Bailey}}{{Bailey}}{1980}]{Bailey1980}
{Bailey} M.~E.,  1980, \mn@doi [\mnras] {10.1093/mnras/191.2.195}, \href
  {http://adsabs.harvard.edu/abs/1980MNRAS.191..195B} {191, 195}

\bibitem[\protect\citeauthoryear{{Bell} et~al.,}{{Bell}
  et~al.}{2011}]{Bell2011}
{Bell} M.~E.,  et~al., 2011, \mn@doi [\mnras]
  {10.1111/j.1365-2966.2010.17692.x}, \href
  {http://adsabs.harvard.edu/abs/2011MNRAS.411..402B} {411, 402}

\bibitem[\protect\citeauthoryear{{Belloni}, {Klein-Wolt}, {M{\'e}ndez}, {van
  der Klis}  \& {van Paradijs}}{{Belloni} et~al.}{2000}]{Belloni2000}
{Belloni} T.,  {Klein-Wolt} M.,  {M{\'e}ndez} M.,  {van der Klis} M.,   {van
  Paradijs} J.,  2000, \aap, \href
  {http://adsabs.harvard.edu/abs/2000A%26A...355..271B} {355, 271}

\bibitem[\protect\citeauthoryear{{Bentz}}{{Bentz}}{2016}]{Bentz2016}
{Bentz} M.~C.,  2016, \mn@doi [Astronomy at High Angular Resolution]
  {10.1007/978-3-319-39739-9_13}, \href
  {http://adsabs.harvard.edu/abs/2016ASSL..439..249B} {439, 249}

\bibitem[\protect\citeauthoryear{{Bianchi}, {Matt}, {Balestra}  \&
  {Perola}}{{Bianchi} et~al.}{2003}]{Bianchi2003}
{Bianchi} S.,  {Matt} G.,  {Balestra} I.,   {Perola} G.~C.,  2003, \mn@doi
  [\aap] {10.1051/0004-6361:20031054}, \href
  {http://adsabs.harvard.edu/abs/2003A%26A...407L..21B} {407, L21}

\bibitem[\protect\citeauthoryear{{Bianchi}, {La Franca}, {Matt}, {Guainazzi},
  {Jimenez Bail{\'o}n}, {Longinotti}, {Nicastro}  \& {Pentericci}}{{Bianchi}
  et~al.}{2008}]{Bianchi2008}
{Bianchi} S.,  {La Franca} F.,  {Matt} G.,  {Guainazzi} M.,  {Jimenez
  Bail{\'o}n} E.,  {Longinotti} A.~L.,  {Nicastro} F.,   {Pentericci} L.,
  2008, \mn@doi [\mnras] {10.1111/j.1745-3933.2008.00521.x}, \href
  {http://adsabs.harvard.edu/abs/2008MNRAS.389L..52B} {389, L52}

\bibitem[\protect\citeauthoryear{{Blaes}}{{Blaes}}{2014}]{Blaes2014}
{Blaes} O.,  2014, \mn@doi [\ssr] {10.1007/s11214-013-9985-6}, \href
  {http://adsabs.harvard.edu/abs/2014SSRv..183...21B} {183, 21}

\bibitem[\protect\citeauthoryear{{Blanchard} et~al.,}{{Blanchard}
  et~al.}{2017}]{Blanchard2017a}
{Blanchard} P.~K.,  et~al., 2017, \mn@doi [\apj] {10.3847/1538-4357/aa77f7},
  \href {http://adsabs.harvard.edu/abs/2017ApJ...843..106B} {843, 106}

\bibitem[\protect\citeauthoryear{{Blank}, {Harnett}  \& {Jones}}{{Blank}
  et~al.}{2005}]{Blank2005}
{Blank} D.~L.,  {Harnett} J.~I.,   {Jones} P.~A.,  2005, \mn@doi [\mnras]
  {10.1111/j.1365-2966.2004.08506.x}, \href
  {http://adsabs.harvard.edu/abs/2005MNRAS.356..734B} {356, 734}

\bibitem[\protect\citeauthoryear{{Bonnerot}, {Rossi}  \& {Lodato}}{{Bonnerot}
  et~al.}{2016}]{Bonnerot2016}
{Bonnerot} C.,  {Rossi} E.~M.,   {Lodato} G.,  2016, \mn@doi [\mnras]
  {10.1093/mnras/stw486}, \href
  {http://adsabs.harvard.edu/abs/2016MNRAS.458.3324B} {458, 3324}

\bibitem[\protect\citeauthoryear{{Burderi}, {King}  \&
  {Szuszkiewicz}}{{Burderi} et~al.}{1998}]{Burderi1998}
{Burderi} L.,  {King} A.~R.,   {Szuszkiewicz} E.,  1998, \mn@doi [\apj]
  {10.1086/306478}, \href {http://adsabs.harvard.edu/abs/1998ApJ...509...85B}
  {509, 85}

\bibitem[\protect\citeauthoryear{{Burrows} et~al.,}{{Burrows}
  et~al.}{2005}]{Burrows2005}
{Burrows} D.~N.,  et~al., 2005, \mn@doi [\ssr] {10.1007/s11214-005-5097-2},
  \href {http://adsabs.harvard.edu/abs/2005SSRv..120..165B} {120, 165}

\bibitem[\protect\citeauthoryear{{Campana}, {Mainetti}, {Colpi}, {Lodato},
  {D'Avanzo}, {Evans}  \& {Moretti}}{{Campana} et~al.}{2015}]{Campana2015}
{Campana} S.,  {Mainetti} D.,  {Colpi} M.,  {Lodato} G.,  {D'Avanzo} P.,
  {Evans} P.~A.,   {Moretti} A.,  2015, \mn@doi [\aap]
  {10.1051/0004-6361/201525965}, \href
  {http://adsabs.harvard.edu/abs/2015A%26A...581A..17C} {581, A17}

\bibitem[\protect\citeauthoryear{{Cannizzo}}{{Cannizzo}}{1994}]{Cannizzo1994}
{Cannizzo} J.~K.,  1994, \mn@doi [\apj] {10.1086/174821}, \href
  {http://adsabs.harvard.edu/abs/1994ApJ...435..389C} {435, 389}

\bibitem[\protect\citeauthoryear{{Cannizzo}, {Chen}  \& {Livio}}{{Cannizzo}
  et~al.}{1995}]{Cannizzo1995}
{Cannizzo} J.~K.,  {Chen} W.,   {Livio} M.,  1995, \mn@doi [\apj]
  {10.1086/176541}, \href {http://adsabs.harvard.edu/abs/1995ApJ...454..880C}
  {454, 880}

\bibitem[\protect\citeauthoryear{{Chen}, {Shrader}  \& {Livio}}{{Chen}
  et~al.}{1997}]{Chen1997}
{Chen} W.,  {Shrader} C.~R.,   {Livio} M.,  1997, \mn@doi [\apj]
  {10.1086/304921}, \href {http://adsabs.harvard.edu/abs/1997ApJ...491..312C}
  {491, 312}

\bibitem[\protect\citeauthoryear{{Czerny}, {Siemiginowska}, {Janiuk},
  {Nikiel-Wroczy{\'n}ski}  \& {Stawarz}}{{Czerny} et~al.}{2009}]{Czerny2009}
{Czerny} B.,  {Siemiginowska} A.,  {Janiuk} A.,  {Nikiel-Wroczy{\'n}ski} B.,
  {Stawarz} {\L}.,  2009, \mn@doi [\apj] {10.1088/0004-637X/698/1/840}, \href
  {http://adsabs.harvard.edu/abs/2009ApJ...698..840C} {698, 840}

\bibitem[\protect\citeauthoryear{{Denney} et~al.,}{{Denney}
  et~al.}{2014}]{Denney2014}
{Denney} K.~D.,  et~al., 2014, \mn@doi [\apj] {10.1088/0004-637X/796/2/134},
  \href {http://adsabs.harvard.edu/abs/2014ApJ...796..134D} {796, 134}

\bibitem[\protect\citeauthoryear{{Done} \& {Gierli{\'n}ski}}{{Done} \&
  {Gierli{\'n}ski}}{2005}]{Done2005}
{Done} C.,  {Gierli{\'n}ski} M.,  2005, \mn@doi [\mnras]
  {10.1111/j.1365-2966.2005.09555.x}, \href
  {http://adsabs.harvard.edu/abs/2005MNRAS.364..208D} {364, 208}

\bibitem[\protect\citeauthoryear{{Done}, {Gierli{\'n}ski}  \& {Kubota}}{{Done}
  et~al.}{2007}]{Done2007}
{Done} C.,  {Gierli{\'n}ski} M.,   {Kubota} A.,  2007, \mn@doi [\aapr]
  {10.1007/s00159-007-0006-1}, \href
  {http://adsabs.harvard.edu/abs/2007A%26ARv..15....1D} {15, 1}

\bibitem[\protect\citeauthoryear{{Drake} et~al.,}{{Drake}
  et~al.}{2011}]{Drake2011}
{Drake} A.~J.,  et~al., 2011, \mn@doi [\apj] {10.1088/0004-637X/735/2/106},
  \href {http://adsabs.harvard.edu/abs/2011ApJ...735..106D} {735, 106}

\bibitem[\protect\citeauthoryear{{Elitzur}, {Ho}  \& {Trump}}{{Elitzur}
  et~al.}{2014}]{Elitzur2014}
{Elitzur} M.,  {Ho} L.~C.,   {Trump} J.~R.,  2014, \mn@doi [\mnras]
  {10.1093/mnras/stt2445}, \href
  {http://adsabs.harvard.edu/abs/2014MNRAS.438.3340E} {438, 3340}

\bibitem[\protect\citeauthoryear{{Emmanoulopoulos}, {Papadakis}, {McHardy},
  {Ar{\'e}valo}, {Calvelo}  \& {Uttley}}{{Emmanoulopoulos}
  et~al.}{2012}]{Emmanoulopoulos2012}
{Emmanoulopoulos} D.,  {Papadakis} I.~E.,  {McHardy} I.~M.,  {Ar{\'e}valo} P.,
  {Calvelo} D.~E.,   {Uttley} P.,  2012, \mn@doi [\mnras]
  {10.1111/j.1365-2966.2012.21316.x}, \href
  {http://adsabs.harvard.edu/abs/2012MNRAS.424.1327E} {424, 1327}

\bibitem[\protect\citeauthoryear{{Emmanoulopoulos}, {Papadakis}, {Nicastro}  \&
  {McHardy}}{{Emmanoulopoulos} et~al.}{2013}]{Emmanoulopoulos2013}
{Emmanoulopoulos} D.,  {Papadakis} I.~E.,  {Nicastro} F.,   {McHardy} I.~M.,
  2013, \mn@doi [\mnras] {10.1093/mnras/sts610}, \href
  {http://adsabs.harvard.edu/abs/2013MNRAS.429.3439E} {429, 3439}

\bibitem[\protect\citeauthoryear{{Evans}}{{Evans}}{1952}]{Evans1952}
{Evans} D.~S.,  1952, \mn@doi [\mnras] {10.1093/mnras/112.6.606}, \href
  {http://adsabs.harvard.edu/abs/1952MNRAS.112..606E} {112, 606}

\bibitem[\protect\citeauthoryear{{Evans} \& {Kochanek}}{{Evans} \&
  {Kochanek}}{1989}]{Evans1989}
{Evans} C.~R.,  {Kochanek} C.~S.,  1989, \mn@doi [\apjl] {10.1086/185567},
  \href {http://adsabs.harvard.edu/abs/1989ApJ...346L..13E} {346, L13}

\bibitem[\protect\citeauthoryear{{Filippenko} \& {Halpern}}{{Filippenko} \&
  {Halpern}}{1984}]{Filippenko1984}
{Filippenko} A.~V.,  {Halpern} J.~P.,  1984, \mn@doi [\apj] {10.1086/162521},
  \href {http://adsabs.harvard.edu/abs/1984ApJ...285..458F} {285, 458}

\bibitem[\protect\citeauthoryear{{Foreman-Mackey}, {Hogg}, {Lang}  \&
  {Goodman}}{{Foreman-Mackey} et~al.}{2013}]{Foreman-Mackey2013}
{Foreman-Mackey} D.,  {Hogg} D.~W.,  {Lang} D.,   {Goodman} J.,  2013, \mn@doi
  [\pasp] {10.1086/670067}, \href
  {http://adsabs.harvard.edu/abs/2013PASP..125..306F} {125, 306}

\bibitem[\protect\citeauthoryear{{Galloway}, {Muno}, {Hartman}, {Psaltis}  \&
  {Chakrabarty}}{{Galloway} et~al.}{2008}]{Galloway2008}
{Galloway} D.~K.,  {Muno} M.~P.,  {Hartman} J.~M.,  {Psaltis} D.,
  {Chakrabarty} D.,  2008, \mn@doi [\apjs] {10.1086/592044}, \href
  {http://adsabs.harvard.edu/abs/2008ApJS..179..360G} {179, 360}

\bibitem[\protect\citeauthoryear{{Gezari} et~al.,}{{Gezari}
  et~al.}{2009}]{Gezari2009}
{Gezari} S.,  et~al., 2009, \mn@doi [\apj] {10.1088/0004-637X/698/2/1367},
  \href {http://adsabs.harvard.edu/abs/2009ApJ...698.1367G} {698, 1367}

\bibitem[\protect\citeauthoryear{{Grupe}, {Komossa}, {Scharw{\"a}chter},
  {Dietrich}, {Leighly}, {Lucy}  \& {Barlow}}{{Grupe} et~al.}{2013}]{Grupe2013}
{Grupe} D.,  {Komossa} S.,  {Scharw{\"a}chter} J.,  {Dietrich} M.,  {Leighly}
  K.~M.,  {Lucy} A.,   {Barlow} B.~N.,  2013, \mn@doi [\aj]
  {10.1088/0004-6256/146/4/78}, \href
  {http://adsabs.harvard.edu/abs/2013AJ....146...78G} {146, 78}

\bibitem[\protect\citeauthoryear{{Grupe}, {Komossa}  \& {Saxton}}{{Grupe}
  et~al.}{2015}]{Grupe2015}
{Grupe} D.,  {Komossa} S.,   {Saxton} R.,  2015, \mn@doi [\apjl]
  {10.1088/2041-8205/803/2/L28}, \href
  {http://adsabs.harvard.edu/abs/2015ApJ...803L..28G} {803, L28}

\bibitem[\protect\citeauthoryear{{Grz{\c e}dzielski}, {Janiuk}, {Czerny}  \&
  {Wu}}{{Grz{\c e}dzielski} et~al.}{2017}]{Grzedzielski2017}
{Grz{\c e}dzielski} M.,  {Janiuk} A.,  {Czerny} B.,   {Wu} Q.,  2017, \mn@doi
  [\aap] {10.1051/0004-6361/201629672}, \href
  {http://adsabs.harvard.edu/abs/2017A%26A...603A.110G} {603, A110}

\bibitem[\protect\citeauthoryear{{Guillochon} \& {Ramirez-Ruiz}}{{Guillochon}
  \& {Ramirez-Ruiz}}{2013}]{Guillochon2013}
{Guillochon} J.,  {Ramirez-Ruiz} E.,  2013, \mn@doi [\apj]
  {10.1088/0004-637X/767/1/25}, \href
  {http://adsabs.harvard.edu/abs/2013ApJ...767...25G} {767, 25}

\bibitem[\protect\citeauthoryear{{Guillochon} \& {Ramirez-Ruiz}}{{Guillochon}
  \& {Ramirez-Ruiz}}{2015}]{Guillochon2015}
{Guillochon} J.,  {Ramirez-Ruiz} E.,  2015, \mn@doi [\apj]
  {10.1088/0004-637X/809/2/166}, \href
  {http://adsabs.harvard.edu/abs/2015ApJ...809..166G} {809, 166}

\bibitem[\protect\citeauthoryear{{Guillochon}, {Manukian}  \&
  {Ramirez-Ruiz}}{{Guillochon} et~al.}{2014}]{Guillochon2014}
{Guillochon} J.,  {Manukian} H.,   {Ramirez-Ruiz} E.,  2014, \mn@doi [\apj]
  {10.1088/0004-637X/783/1/23}, \href
  {http://adsabs.harvard.edu/abs/2014ApJ...783...23G} {783, 23}

\bibitem[\protect\citeauthoryear{{G{\"u}ltekin} et~al.,}{{G{\"u}ltekin}
  et~al.}{2009}]{Gultekin2009}
{G{\"u}ltekin} K.,  et~al., 2009, \mn@doi [\apj] {10.1088/0004-637X/698/1/198},
  \href {http://adsabs.harvard.edu/abs/2009ApJ...698..198G} {698, 198}

\bibitem[\protect\citeauthoryear{{Halpern} \& {Filippenko}}{{Halpern} \&
  {Filippenko}}{1984}]{Halpern1984}
{Halpern} J.~P.,  {Filippenko} A.~V.,  1984, \mn@doi [\apj] {10.1086/162522},
  \href {http://adsabs.harvard.edu/abs/1984ApJ...285..475H} {285, 475}

\bibitem[\protect\citeauthoryear{{Hameury}, {Viallet}  \& {Lasota}}{{Hameury}
  et~al.}{2009}]{Hameury2009}
{Hameury} J.-M.,  {Viallet} M.,   {Lasota} J.-P.,  2009, \mn@doi [\aap]
  {10.1051/0004-6361/200810928}, \href
  {http://adsabs.harvard.edu/abs/2009A%26A...496..413H} {496, 413}

\bibitem[\protect\citeauthoryear{{Hancock}, {Tingay}, {Sadler}, {Phillips}  \&
  {Deller}}{{Hancock} et~al.}{2009}]{Hancock2009}
{Hancock} P.~J.,  {Tingay} S.~J.,  {Sadler} E.~M.,  {Phillips} C.,   {Deller}
  A.~T.,  2009, \mn@doi [\mnras] {10.1111/j.1365-2966.2009.15055.x}, \href
  {http://adsabs.harvard.edu/abs/2009MNRAS.397.2030H} {397, 2030}

\bibitem[\protect\citeauthoryear{{Hayasaki}, {Stone}  \& {Loeb}}{{Hayasaki}
  et~al.}{2016}]{Hayasaki2016}
{Hayasaki} K.,  {Stone} N.,   {Loeb} A.,  2016, \mn@doi [\mnras]
  {10.1093/mnras/stw1387}, \href
  {http://adsabs.harvard.edu/abs/2016MNRAS.461.3760H} {461, 3760}

\bibitem[\protect\citeauthoryear{{Hirose}, {Krolik}  \& {Blaes}}{{Hirose}
  et~al.}{2009}]{Hirose2009}
{Hirose} S.,  {Krolik} J.~H.,   {Blaes} O.,  2009, \mn@doi [\apj]
  {10.1088/0004-637X/691/1/16}, \href
  {http://adsabs.harvard.edu/abs/2009ApJ...691...16H} {691, 16}

\bibitem[\protect\citeauthoryear{{Ho}}{{Ho}}{2008}]{Ho2008}
{Ho} L.~C.,  2008, \mn@doi [\araa] {10.1146/annurev.astro.45.051806.110546},
  \href {http://adsabs.harvard.edu/abs/2008ARA%26A..46..475H} {46, 475}

\bibitem[\protect\citeauthoryear{{Janiuk} \& {Czerny}}{{Janiuk} \&
  {Czerny}}{2011}]{Janiuk2011}
{Janiuk} A.,  {Czerny} B.,  2011, \mn@doi [\mnras]
  {10.1111/j.1365-2966.2011.18544.x}, \href
  {http://adsabs.harvard.edu/abs/2011MNRAS.414.2186J} {414, 2186}

\bibitem[\protect\citeauthoryear{{Janiuk}, {Czerny}  \&
  {Siemiginowska}}{{Janiuk} et~al.}{2002}]{Janiuk2002}
{Janiuk} A.,  {Czerny} B.,   {Siemiginowska} A.,  2002, \mn@doi [\apj]
  {10.1086/341804}, \href {http://adsabs.harvard.edu/abs/2002ApJ...576..908J}
  {576, 908}

\bibitem[\protect\citeauthoryear{{Janiuk}, {Czerny}, {Siemiginowska}  \&
  {Szczerba}}{{Janiuk} et~al.}{2004}]{Janiuk2004}
{Janiuk} A.,  {Czerny} B.,  {Siemiginowska} A.,   {Szczerba} R.,  2004, \mn@doi
  [\apj] {10.1086/381159}, \href
  {http://adsabs.harvard.edu/abs/2004ApJ...602..595J} {602, 595}

\bibitem[\protect\citeauthoryear{{Jiang}, {Stone}  \& {Davis}}{{Jiang}
  et~al.}{2013}]{Jiang2013}
{Jiang} Y.-F.,  {Stone} J.~M.,   {Davis} S.~W.,  2013, \mn@doi [\apj]
  {10.1088/0004-637X/778/1/65}, \href
  {http://adsabs.harvard.edu/abs/2013ApJ...778...65J} {778, 65}

\bibitem[\protect\citeauthoryear{{Karas} \& {{\v S}ubr}}{{Karas} \& {{\v
  S}ubr}}{2007}]{Karas2007}
{Karas} V.,  {{\v S}ubr} L.,  2007, \mn@doi [\aap]
  {10.1051/0004-6361:20066068}, \href
  {http://adsabs.harvard.edu/abs/2007A%26A...470...11K} {470, 11}

\bibitem[\protect\citeauthoryear{{Kathirgamaraju}, {Barniol Duran}  \&
  {Giannios}}{{Kathirgamaraju} et~al.}{2017}]{Kathirgamaraju2017}
{Kathirgamaraju} A.,  {Barniol Duran} R.,   {Giannios} D.,  2017, \mn@doi
  [\mnras] {10.1093/mnras/stx846}, \href
  {http://adsabs.harvard.edu/abs/2017MNRAS.469..314K} {469, 314}

\bibitem[\protect\citeauthoryear{{Kelly}, {Bechtold}  \&
  {Siemiginowska}}{{Kelly} et~al.}{2009}]{Kelly2009}
{Kelly} B.~C.,  {Bechtold} J.,   {Siemiginowska} A.,  2009, \mn@doi [\apj]
  {10.1088/0004-637X/698/1/895}, \href
  {http://adsabs.harvard.edu/abs/2009ApJ...698..895K} {698, 895}

\bibitem[\protect\citeauthoryear{{Kelly}, {Sobolewska}  \&
  {Siemiginowska}}{{Kelly} et~al.}{2011}]{Kelly2011}
{Kelly} B.~C.,  {Sobolewska} M.,   {Siemiginowska} A.,  2011, \mn@doi [\apj]
  {10.1088/0004-637X/730/1/52}, \href
  {http://adsabs.harvard.edu/abs/2011ApJ...730...52K} {730, 52}

\bibitem[\protect\citeauthoryear{{Kennedy}, {Meiron}, {Shukirgaliyev},
  {Panamarev}, {Berczik}, {Just}  \& {Spurzem}}{{Kennedy}
  et~al.}{2016}]{Kennedy2016}
{Kennedy} G.~F.,  {Meiron} Y.,  {Shukirgaliyev} B.,  {Panamarev} T.,  {Berczik}
  P.,  {Just} A.,   {Spurzem} R.,  2016, \mn@doi [\mnras]
  {10.1093/mnras/stw908}, \href
  {http://adsabs.harvard.edu/abs/2016MNRAS.460..240K} {460, 240}

\bibitem[\protect\citeauthoryear{{King} \& {Ritter}}{{King} \&
  {Ritter}}{1998}]{King1998}
{King} A.~R.,  {Ritter} H.,  1998, \mn@doi [\mnras]
  {10.1046/j.1365-8711.1998.01295.x}, \href
  {http://adsabs.harvard.edu/abs/1998MNRAS.293L..42K} {293, L42}

\bibitem[\protect\citeauthoryear{{Komossa}}{{Komossa}}{2015}]{Komossa2015a}
{Komossa} S.,  2015, \mn@doi [Journal of High Energy Astrophysics]
  {10.1016/j.jheap.2015.04.006}, \href
  {http://adsabs.harvard.edu/abs/2015JHEAp...7..148K} {7, 148}

\bibitem[\protect\citeauthoryear{{LaMassa} et~al.,}{{LaMassa}
  et~al.}{2015}]{LaMassa2015}
{LaMassa} S.~M.,  et~al., 2015, \mn@doi [\apj] {10.1088/0004-637X/800/2/144},
  \href {http://adsabs.harvard.edu/abs/2015ApJ...800..144L} {800, 144}

\bibitem[\protect\citeauthoryear{{Lasota}}{{Lasota}}{2001}]{Lasota2001}
{Lasota} J.-P.,  2001, \mn@doi [\nar] {10.1016/S1387-6473(01)00112-9}, \href
  {http://adsabs.harvard.edu/abs/2001NewAR..45..449L} {45, 449}

\bibitem[\protect\citeauthoryear{{Lasota}}{{Lasota}}{2012}]{Lasota2012}
{Lasota} J.-P.,  2012, \memsai, \href
  {http://adsabs.harvard.edu/abs/2012MmSAI..83..469L} {83, 469}

\bibitem[\protect\citeauthoryear{{Lasota}}{{Lasota}}{2016}]{Lasota2016}
{Lasota} J.-P.,  2016, in {Bambi} C.,  ed.,  Astrophysics and Space Science
  Library Vol. 440, Astrophysics of Black Holes: From Fundamental Aspects to
  Latest Developments. p.~1 (\mn@eprint {arXiv} {1505.02172}),
  \mn@doi{10.1007/978-3-662-52859-4_1}

\bibitem[\protect\citeauthoryear{{Lasota}, {Alexander}, {Dubus}, {Barret},
  {Farrell}, {Gehrels}, {Godet}  \& {Webb}}{{Lasota} et~al.}{2011}]{Lasota2011}
{Lasota} J.-P.,  {Alexander} T.,  {Dubus} G.,  {Barret} D.,  {Farrell} S.~A.,
  {Gehrels} N.,  {Godet} O.,   {Webb} N.~A.,  2011, \mn@doi [\apj]
  {10.1088/0004-637X/735/2/89}, \href
  {http://adsabs.harvard.edu/abs/2011ApJ...735...89L} {735, 89}

\bibitem[\protect\citeauthoryear{{Lightman} \& {Eardley}}{{Lightman} \&
  {Eardley}}{1974}]{Lightman1974}
{Lightman} A.~P.,  {Eardley} D.~M.,  1974, \mn@doi [\apjl] {10.1086/181377},
  \href {http://adsabs.harvard.edu/abs/1974ApJ...187L...1L} {187, L1}

\bibitem[\protect\citeauthoryear{{Lin} \& {Shields}}{{Lin} \&
  {Shields}}{1986}]{Lin1986}
{Lin} D.~N.~C.,  {Shields} G.~A.,  1986, \mn@doi [\apj] {10.1086/164225}, \href
  {http://adsabs.harvard.edu/abs/1986ApJ...305...28L} {305, 28}

\bibitem[\protect\citeauthoryear{{Lin} et~al.,}{{Lin} et~al.}{2017a}]{Lin2017a}
{Lin} D.,  et~al., 2017a, \mn@doi [Nature Astronomy] {10.1038/s41550-016-0033},
  \href {http://adsabs.harvard.edu/abs/2017NatAs...1E..33L} {1, 0033}

\bibitem[\protect\citeauthoryear{{Lin}, {Godet}, {Ho}, {Barret}, {Webb}  \&
  {Irwin}}{{Lin} et~al.}{2017b}]{Lin2017}
{Lin} D.,  {Godet} O.,  {Ho} L.~C.,  {Barret} D.,  {Webb} N.~A.,   {Irwin}
  J.~A.,  2017b, \mn@doi [\mnras] {10.1093/mnras/stx489}, \href
  {http://adsabs.harvard.edu/abs/2017MNRAS.468..783L} {468, 783}

\bibitem[\protect\citeauthoryear{{Lobban}, {Reeves}, {Porquet}, {Braito},
  {Markowitz}, {Miller}  \& {Turner}}{{Lobban} et~al.}{2010}]{Lobban2010}
{Lobban} A.~P.,  {Reeves} J.~N.,  {Porquet} D.,  {Braito} V.,  {Markowitz} A.,
  {Miller} L.,   {Turner} T.~J.,  2010, \mn@doi [\mnras]
  {10.1111/j.1365-2966.2010.17143.x}, \href
  {http://adsabs.harvard.edu/abs/2010MNRAS.408..551L} {408, 551}

\bibitem[\protect\citeauthoryear{{MacLeod}, {Guillochon}  \&
  {Ramirez-Ruiz}}{{MacLeod} et~al.}{2012}]{MacLeod2012}
{MacLeod} M.,  {Guillochon} J.,   {Ramirez-Ruiz} E.,  2012, \mn@doi [\apj]
  {10.1088/0004-637X/757/2/134}, \href
  {http://adsabs.harvard.edu/abs/2012ApJ...757..134M} {757, 134}

\bibitem[\protect\citeauthoryear{{Marshall}, {Boldt}, {Holt}, {Mushotzky},
  {Rothschild}, {Serlemitsos}  \& {Pravdo}}{{Marshall}
  et~al.}{1979}]{Marshall1979}
{Marshall} F.~E.,  {Boldt} E.~A.,  {Holt} S.~S.,  {Mushotzky} R.~F.,
  {Rothschild} R.~E.,  {Serlemitsos} P.~J.,   {Pravdo} S.~H.,  1979, \mn@doi
  [\apjs] {10.1086/190600}, \href
  {http://adsabs.harvard.edu/abs/1979ApJS...40..657M} {40, 657}

\bibitem[\protect\citeauthoryear{{McHardy}}{{McHardy}}{2010}]{McHardy2010}
{McHardy} I.,  2010, in {Belloni} T.,  ed.,  Lecture Notes in Physics, Berlin
  Springer Verlag Vol. 794, Lecture Notes in Physics, Berlin Springer Verlag.
  p.~203 (\mn@eprint {arXiv} {0909.2579}), \mn@doi{10.1007/978-3-540-76937-8_8}

\bibitem[\protect\citeauthoryear{{McHardy}, {Papadakis}, {Uttley}, {Page}  \&
  {Mason}}{{McHardy} et~al.}{2004}]{McHardy2004}
{McHardy} I.~M.,  {Papadakis} I.~E.,  {Uttley} P.,  {Page} M.~J.,   {Mason}
  K.~O.,  2004, \mn@doi [\mnras] {10.1111/j.1365-2966.2004.07376.x}, \href
  {http://adsabs.harvard.edu/abs/2004MNRAS.348..783M} {348, 783}

\bibitem[\protect\citeauthoryear{{McHardy}, {Koerding}, {Knigge}, {Uttley}  \&
  {Fender}}{{McHardy} et~al.}{2006}]{McHardy2006}
{McHardy} I.~M.,  {Koerding} E.,  {Knigge} C.,  {Uttley} P.,   {Fender} R.~P.,
  2006, \mn@doi [\nat] {10.1038/nature05389}, \href
  {http://adsabs.harvard.edu/abs/2006Natur.444..730M} {444, 730}

\bibitem[\protect\citeauthoryear{{Merloni} et~al.,}{{Merloni}
  et~al.}{2015}]{Merloni2015}
{Merloni} A.,  et~al., 2015, \mn@doi [\mnras] {10.1093/mnras/stv1095}, \href
  {http://adsabs.harvard.edu/abs/2015MNRAS.452...69M} {452, 69}

\bibitem[\protect\citeauthoryear{{Mishra}, {Fragile}, {Johnson}  \&
  {Klu{\'z}niak}}{{Mishra} et~al.}{2016}]{Mishra2016}
{Mishra} B.,  {Fragile} P.~C.,  {Johnson} L.~C.,   {Klu{\'z}niak} W.,  2016,
  \mn@doi [\mnras] {10.1093/mnras/stw2245}, \href
  {http://adsabs.harvard.edu/abs/2016MNRAS.463.3437M} {463, 3437}

\bibitem[\protect\citeauthoryear{{Montesinos Armijo} \& {de Freitas
  Pacheco}}{{Montesinos Armijo} \& {de Freitas
  Pacheco}}{2011}]{Montesinos-Armijo2011}
{Montesinos Armijo} M.,  {de Freitas Pacheco} J.~A.,  2011, \mn@doi [\apj]
  {10.1088/0004-637X/736/2/126}, \href
  {http://adsabs.harvard.edu/abs/2011ApJ...736..126M} {736, 126}

\bibitem[\protect\citeauthoryear{{Nandra} \& {Pounds}}{{Nandra} \&
  {Pounds}}{1994}]{Nandra1994}
{Nandra} K.,  {Pounds} K.~A.,  1994, \mn@doi [\mnras]
  {10.1093/mnras/268.2.405}, \href
  {http://adsabs.harvard.edu/abs/1994MNRAS.268..405N} {268, 405}

\bibitem[\protect\citeauthoryear{{Netzer}}{{Netzer}}{2008}]{Netzer2008}
{Netzer} H.,  2008, \mn@doi [\nar] {10.1016/j.newar.2008.06.009}, \href
  {http://adsabs.harvard.edu/abs/2008NewAR..52..257N} {52, 257}

\bibitem[\protect\citeauthoryear{{Papadakis}}{{Papadakis}}{2004}]{Papadakis2004}
{Papadakis} I.~E.,  2004, \mn@doi [\mnras] {10.1111/j.1365-2966.2004.07351.x},
  \href {http://adsabs.harvard.edu/abs/2004MNRAS.348..207P} {348, 207}

\bibitem[\protect\citeauthoryear{{Peng}, {Yin}, {Bi}, {Zhao}, {Fang}, {Bao}  \&
  {Ma}}{{Peng} et~al.}{2010}]{Peng2010}
{Peng} Z.~Y.,  {Yin} Y.,  {Bi} X.~W.,  {Zhao} X.~H.,  {Fang} L.~M.,  {Bao}
  Y.~Y.,   {Ma} L.,  2010, \mn@doi [\apj] {10.1088/0004-637X/718/2/894}, \href
  {http://adsabs.harvard.edu/abs/2010ApJ...718..894P} {718, 894}

\bibitem[\protect\citeauthoryear{{Peterson}}{{Peterson}}{2001}]{Peterson2001}
{Peterson} B.~M.,  2001, in {Aretxaga} I.,  {Kunth} D.,   {M{\'u}jica} R.,
  eds, Advanced Lectures on the Starburst-AGN. p.~3 (\mn@eprint {}
  {astro-ph/0109495}), \mn@doi{10.1142/9789812811318_0002}

\bibitem[\protect\citeauthoryear{{Peterson} \& {Bentz}}{{Peterson} \&
  {Bentz}}{2006}]{Peterson2006}
{Peterson} B.~M.,  {Bentz} M.~C.,  2006, \mn@doi [\nar]
  {10.1016/j.newar.2006.06.062}, \href
  {http://adsabs.harvard.edu/abs/2006NewAR..50..796P} {50, 796}

\bibitem[\protect\citeauthoryear{{Phillips}}{{Phillips}}{1979}]{Phillips1979}
{Phillips} M.~M.,  1979, \mn@doi [\apjl] {10.1086/182881}, \href
  {http://adsabs.harvard.edu/abs/1979ApJ...227L.121P} {227, L121}

\bibitem[\protect\citeauthoryear{{Phinney}}{{Phinney}}{1989}]{Phinney1989}
{Phinney} E.~S.,  1989, in {Morris} M.,  ed.,  IAU Symposium Vol. 136, The
  Center of the Galaxy. p.~543

\bibitem[\protect\citeauthoryear{{Piccinotti}, {Mushotzky}, {Boldt}, {Holt},
  {Marshall}, {Serlemitsos}  \& {Shafer}}{{Piccinotti}
  et~al.}{1982}]{Piccinotti1982}
{Piccinotti} G.,  {Mushotzky} R.~F.,  {Boldt} E.~A.,  {Holt} S.~S.,  {Marshall}
  F.~E.,  {Serlemitsos} P.~J.,   {Shafer} R.~A.,  1982, \mn@doi [\apj]
  {10.1086/159651}, \href {http://adsabs.harvard.edu/abs/1982ApJ...253..485P}
  {253, 485}

\bibitem[\protect\citeauthoryear{{Planck Collaboration} et~al.,}{{Planck
  Collaboration} et~al.}{2016}]{Planck-Collaboration2016}
{Planck Collaboration} et~al., 2016, \mn@doi [\aap]
  {10.1051/0004-6361/201525830}, \href
  {http://adsabs.harvard.edu/abs/2016A%26A...594A..13P} {594, A13}

\bibitem[\protect\citeauthoryear{{Powell}, {Haswell}  \& {Falanga}}{{Powell}
  et~al.}{2007}]{Powell2007}
{Powell} C.~R.,  {Haswell} C.~A.,   {Falanga} M.,  2007, \mn@doi [\mnras]
  {10.1111/j.1365-2966.2006.11144.x}, \href
  {http://adsabs.harvard.edu/abs/2007MNRAS.374..466P} {374, 466}

\bibitem[\protect\citeauthoryear{{Prieto}, {Reunanen}  \&
  {Kotilainen}}{{Prieto} et~al.}{2002}]{Prieto2002}
{Prieto} M.~A.,  {Reunanen} J.,   {Kotilainen} J.~K.,  2002, \mn@doi [\apjl]
  {10.1086/341201}, \href {http://adsabs.harvard.edu/abs/2002ApJ...571L...7P}
  {571, L7}

\bibitem[\protect\citeauthoryear{{Rees}}{{Rees}}{1988}]{Rees1988}
{Rees} M.~J.,  1988, \mn@doi [\nat] {10.1038/333523a0}, \href
  {http://adsabs.harvard.edu/abs/1988Natur.333..523R} {333, 523}

\bibitem[\protect\citeauthoryear{{Runnoe} et~al.,}{{Runnoe}
  et~al.}{2016}]{Runnoe2016}
{Runnoe} J.~C.,  et~al., 2016, \mn@doi [\mnras] {10.1093/mnras/stv2385}, \href
  {http://adsabs.harvard.edu/abs/2016MNRAS.455.1691R} {455, 1691}

\bibitem[\protect\citeauthoryear{{Ruschel-Dutra}, {Pastoriza}, {Riffel},
  {Sales}  \& {Winge}}{{Ruschel-Dutra} et~al.}{2014}]{Ruschel-Dutra2014}
{Ruschel-Dutra} D.,  {Pastoriza} M.,  {Riffel} R.,  {Sales} D.~A.,   {Winge}
  C.,  2014, \mn@doi [\mnras] {10.1093/mnras/stt2448}, \href
  {http://adsabs.harvard.edu/abs/2014MNRAS.438.3434R} {438, 3434}

\bibitem[\protect\citeauthoryear{{Sadler}}{{Sadler}}{1984}]{Sadler1984}
{Sadler} E.~M.,  1984, \mn@doi [\aj] {10.1086/113483}, \href
  {http://adsabs.harvard.edu/abs/1984AJ.....89...53S} {89}

\bibitem[\protect\citeauthoryear{{Saxton}, {Motta}, {Komossa}  \&
  {Read}}{{Saxton} et~al.}{2015}]{Saxton2015}
{Saxton} R.~D.,  {Motta} S.~E.,  {Komossa} S.,   {Read} A.~M.,  2015, \mn@doi
  [\mnras] {10.1093/mnras/stv2160}, \href
  {http://adsabs.harvard.edu/abs/2015MNRAS.454.2798S} {454, 2798}

\bibitem[\protect\citeauthoryear{{S{\c a}dowski}}{{S{\c
  a}dowski}}{2016}]{Sc-adowski2016}
{S{\c a}dowski} A.,  2016, \mn@doi [\mnras] {10.1093/mnras/stw913}, \href
  {http://adsabs.harvard.edu/abs/2016MNRAS.459.4397S} {459, 4397}

\bibitem[\protect\citeauthoryear{{Schnorr-M{\"u}ller}, {Storchi-Bergmann},
  {Nagar}  \& {Ferrari}}{{Schnorr-M{\"u}ller}
  et~al.}{2014}]{Schnorr-Muller2014}
{Schnorr-M{\"u}ller} A.,  {Storchi-Bergmann} T.,  {Nagar} N.~M.,   {Ferrari}
  F.,  2014, \mn@doi [\mnras] {10.1093/mnras/stt2440}, \href
  {http://adsabs.harvard.edu/abs/2014MNRAS.438.3322S} {438, 3322}

\bibitem[\protect\citeauthoryear{{Shakura} \& {Sunyaev}}{{Shakura} \&
  {Sunyaev}}{1973}]{Shakura1973}
{Shakura} N.~I.,  {Sunyaev} R.~A.,  1973, \aap, \href
  {http://adsabs.harvard.edu/abs/1973A%26A....24..337S} {24, 337}

\bibitem[\protect\citeauthoryear{{Shappee} et~al.,}{{Shappee}
  et~al.}{2014}]{Shappee2014}
{Shappee} B.~J.,  et~al., 2014, \mn@doi [\apj] {10.1088/0004-637X/788/1/48},
  \href {http://adsabs.harvard.edu/abs/2014ApJ...788...48S} {788, 48}

\bibitem[\protect\citeauthoryear{{Shields} \& {Wheeler}}{{Shields} \&
  {Wheeler}}{1978}]{Shields1978}
{Shields} G.~A.,  {Wheeler} J.~C.,  1978, \mn@doi [\apj] {10.1086/156185},
  \href {http://adsabs.harvard.edu/abs/1978ApJ...222..667S} {222, 667}

\bibitem[\protect\citeauthoryear{{Shiokawa}, {Krolik}, {Cheng}, {Piran}  \&
  {Noble}}{{Shiokawa} et~al.}{2015}]{Shiokawa2015}
{Shiokawa} H.,  {Krolik} J.~H.,  {Cheng} R.~M.,  {Piran} T.,   {Noble} S.~C.,
  2015, \mn@doi [\apj] {10.1088/0004-637X/804/2/85}, \href
  {http://adsabs.harvard.edu/abs/2015ApJ...804...85S} {804, 85}

\bibitem[\protect\citeauthoryear{{Siemiginowska}, {Czerny}  \&
  {Kostyunin}}{{Siemiginowska} et~al.}{1996}]{Siemiginowska1996}
{Siemiginowska} A.,  {Czerny} B.,   {Kostyunin} V.,  1996, \mn@doi [\apj]
  {10.1086/176831}, \href {http://adsabs.harvard.edu/abs/1996ApJ...458..491S}
  {458, 491}

\bibitem[\protect\citeauthoryear{{Starling}, {Page}, {Branduardi-Raymont},
  {Breeveld}, {Soria}  \& {Wu}}{{Starling} et~al.}{2005}]{Starling2005}
{Starling} R.~L.~C.,  {Page} M.~J.,  {Branduardi-Raymont} G.,  {Breeveld}
  A.~A.,  {Soria} R.,   {Wu} K.,  2005, \mn@doi [\mnras]
  {10.1111/j.1365-2966.2004.08493.x}, \href
  {http://adsabs.harvard.edu/abs/2005MNRAS.356..727S} {356, 727}

\bibitem[\protect\citeauthoryear{{Storchi-Bergmann}, {Rodriguez-Ardila},
  {Schmitt}, {Wilson}  \& {Baldwin}}{{Storchi-Bergmann}
  et~al.}{1996}]{Storchi-Bergmann1996}
{Storchi-Bergmann} T.,  {Rodriguez-Ardila} A.,  {Schmitt} H.~R.,  {Wilson}
  A.~S.,   {Baldwin} J.~A.,  1996, \mn@doi [\apj] {10.1086/178043}, \href
  {http://adsabs.harvard.edu/abs/1996ApJ...472...83S} {472, 83}

\bibitem[\protect\citeauthoryear{{Strotjohann}, {Saxton}, {Starling}, {Esquej},
  {Read}, {Evans}  \& {Miniutti}}{{Strotjohann} et~al.}{2016}]{Strotjohann2016}
{Strotjohann} N.~L.,  {Saxton} R.~D.,  {Starling} R.~L.~C.,  {Esquej} P.,
  {Read} A.~M.,  {Evans} P.~A.,   {Miniutti} G.,  2016, \mn@doi [\aap]
  {10.1051/0004-6361/201628241}, \href
  {http://adsabs.harvard.edu/abs/2016A%26A...592A..74S} {592, A74}

\bibitem[\protect\citeauthoryear{{Sun}, {Gu}, {Yan}, {Wu}  \& {Liu}}{{Sun}
  et~al.}{2016}]{Sun2016}
{Sun} M.,  {Gu} W.-M.,  {Yan} Z.,  {Wu} Q.,   {Liu} T.,  2016, \mn@doi [\mnras]
  {10.1093/mnrasl/slw159}, \href
  {http://adsabs.harvard.edu/abs/2016MNRAS.463L..99S} {463, L99}

\bibitem[\protect\citeauthoryear{{Tout}, {Pols}, {Eggleton}  \& {Han}}{{Tout}
  et~al.}{1996}]{Tout1996}
{Tout} C.~A.,  {Pols} O.~R.,  {Eggleton} P.~P.,   {Han} Z.,  1996, \mn@doi
  [\mnras] {10.1093/mnras/281.1.257}, \href
  {http://adsabs.harvard.edu/abs/1996MNRAS.281..257T} {281, 257}

\bibitem[\protect\citeauthoryear{{Turner} \& {Pounds}}{{Turner} \&
  {Pounds}}{1989}]{Turner1989}
{Turner} T.~J.,  {Pounds} K.~A.,  1989, \mn@doi [\mnras]
  {10.1093/mnras/240.4.833}, \href
  {http://adsabs.harvard.edu/abs/1989MNRAS.240..833T} {240, 833}

\bibitem[\protect\citeauthoryear{{Turner}, {Nandra}, {Turcan}  \&
  {George}}{{Turner} et~al.}{2001}]{Turner2001}
{Turner} T.~J.,  {Nandra} K.,  {Turcan} D.,   {George} I.~M.,  2001, \mn@doi
  [X-ray Astronomy: Stellar Endpoints, AGN, and the Diffuse X-ray Background]
  {10.1063/1.1434792}, \href
  {http://adsabs.harvard.edu/abs/2001AIPC..599..991T} {599, 991}

\bibitem[\protect\citeauthoryear{{Ulmer}}{{Ulmer}}{1999}]{Ulmer1999}
{Ulmer} A.,  1999, \mn@doi [\apj] {10.1086/306909}, \href
  {http://adsabs.harvard.edu/abs/1999ApJ...514..180U} {514, 180}

\bibitem[\protect\citeauthoryear{{Ulrich}, {Maraschi}  \& {Urry}}{{Ulrich}
  et~al.}{1997}]{Ulrich1997}
{Ulrich} M.-H.,  {Maraschi} L.,   {Urry} C.~M.,  1997, \mn@doi [\araa]
  {10.1146/annurev.astro.35.1.445}, \href
  {http://adsabs.harvard.edu/abs/1997ARA%26A..35..445U} {35, 445}

\bibitem[\protect\citeauthoryear{{Ursini} et~al.,}{{Ursini}
  et~al.}{2015}]{Ursini2015}
{Ursini} F.,  et~al., 2015, \mn@doi [\mnras] {10.1093/mnras/stv1527}, \href
  {http://adsabs.harvard.edu/abs/2015MNRAS.452.3266U} {452, 3266}

\bibitem[\protect\citeauthoryear{{Uttley}, {McHardy}  \& {Vaughan}}{{Uttley}
  et~al.}{2005}]{Uttley2005}
{Uttley} P.,  {McHardy} I.~M.,   {Vaughan} S.,  2005, \mn@doi [\mnras]
  {10.1111/j.1365-2966.2005.08886.x}, \href
  {http://adsabs.harvard.edu/abs/2005MNRAS.359..345U} {359, 345}

\bibitem[\protect\citeauthoryear{{Valtonen} et~al.,}{{Valtonen}
  et~al.}{2008}]{Valtonen2008}
{Valtonen} M.~J.,  et~al., 2008, \mn@doi [\nat] {10.1038/nature06896}, \href
  {http://adsabs.harvard.edu/abs/2008Natur.452..851V} {452, 851}

\bibitem[\protect\citeauthoryear{{Wright}}{{Wright}}{1974}]{Wright1974}
{Wright} A.~E.,  1974, \mn@doi [\mnras] {10.1093/mnras/167.2.273}, \href
  {http://adsabs.harvard.edu/abs/1974MNRAS.167..273W} {167, 273}

\bibitem[\protect\citeauthoryear{{Wright}, {Savage}  \& {Bolton}}{{Wright}
  et~al.}{1977}]{Wright1977}
{Wright} A.~E.,  {Savage} A.,   {Bolton} J.~G.,  1977, Australian Journal of
  Physics Astrophysical Supplement, \href
  {http://adsabs.harvard.edu/abs/1977AuJPA..41....1W} {41, 1}

\bibitem[\protect\citeauthoryear{{Wu}}{{Wu}}{2009}]{Wu2009}
{Wu} Q.,  2009, \mn@doi [\apjl] {10.1088/0004-637X/701/2/L95}, \href
  {http://adsabs.harvard.edu/abs/2009ApJ...701L..95W} {701, L95}

\bibitem[\protect\citeauthoryear{{Wu} et~al.,}{{Wu} et~al.}{2016}]{Wu2016}
{Wu} Q.,  et~al., 2016, \mn@doi [\apj] {10.3847/1538-4357/833/1/79}, \href
  {http://adsabs.harvard.edu/abs/2016ApJ...833...79W} {833, 79}

\bibitem[\protect\citeauthoryear{{Xie}, {Zdziarski}, {Ma}  \& {Yang}}{{Xie}
  et~al.}{2016}]{Xie2016}
{Xie} F.-G.,  {Zdziarski} A.~A.,  {Ma} R.,   {Yang} Q.-X.,  2016, \mn@doi
  [\mnras] {10.1093/mnras/stw2132}, \href
  {http://adsabs.harvard.edu/abs/2016MNRAS.463.2287X} {463, 2287}

\bibitem[\protect\citeauthoryear{{Yan} \& {Yu}}{{Yan} \& {Yu}}{2015}]{Yan2015c}
{Yan} Z.,  {Yu} W.,  2015, \mn@doi [\apj] {10.1088/0004-637X/805/2/87}, \href
  {http://adsabs.harvard.edu/abs/2015ApJ...805...87Y} {805, 87}

\bibitem[\protect\citeauthoryear{{Yuan} \& {Narayan}}{{Yuan} \&
  {Narayan}}{2014}]{Yuan2014}
{Yuan} F.,  {Narayan} R.,  2014, \mn@doi [\araa]
  {10.1146/annurev-astro-082812-141003}, \href
  {http://adsabs.harvard.edu/abs/2014ARA%26A..52..529Y} {52, 529}

\bibitem[\protect\citeauthoryear{{Zhang}, {Fan}, {Dyks}, {Kobayashi},
  {M{\'e}sz{\'a}ros}, {Burrows}, {Nousek}  \& {Gehrels}}{{Zhang}
  et~al.}{2006}]{Zhang2006}
{Zhang} B.,  {Fan} Y.~Z.,  {Dyks} J.,  {Kobayashi} S.,  {M{\'e}sz{\'a}ros} P.,
  {Burrows} D.~N.,  {Nousek} J.~A.,   {Gehrels} N.,  2006, \mn@doi [\apj]
  {10.1086/500723}, \href {http://adsabs.harvard.edu/abs/2006ApJ...642..354Z}
  {642, 354}

\makeatother
\end{thebibliography}
	
	\input{NGC7213.bbl}

\end{document}